\documentclass[aps,prl,twocolumn,noshowpacs,superscriptaddress,groupedaddress]{revtex4}  
\usepackage{graphicx}  
\usepackage{dcolumn}   
\usepackage{bm}        
\usepackage{amssymb}   
\usepackage{color}
\usepackage{amsmath}
\begin{document}

\title{Inverse Statistical Physics of Protein Sequences: A Key Issues
  Review} 
\author{Simona Cocco}
\affiliation{Laboratoire de Physique Statistique de l'Ecole Normale
  Sup\'erieure-UMR 8549, CNRS and PSL Research, Sorbonne Universit\'es
  UPMC, Paris, France} 
\author{Christoph Feinauer}
\affiliation{Sorbonne Universit\'{e}s, UPMC, Institut de Biologie
  Paris-Seine, CNRS, Laboratoire de Biologie Computationnelle et
  Quantitative UMR 7238, Paris, France} 
\author{Matteo Figliuzzi}
\affiliation{Sorbonne Universit\'{e}s, UPMC, Institut de Biologie
  Paris-Seine, CNRS, Laboratoire de Biologie Computationnelle et
  Quantitative UMR 7238, Paris, France} 
\author{R\'emi Monasson}
\affiliation{Laboratoire de Physique Th\'eorique de l'Ecole Normale
  Sup\'erieure-UMR 8549, CNRS and PSL Research, Sorbonne Universit\'e
  UPMC, Paris, France} 
\author{Martin Weigt}
\affiliation{Sorbonne Universit\'{e}s, UPMC, Institut de Biologie
  Paris-Seine, CNRS, Laboratoire de Biologie Computationnelle et
  Quantitative UMR 7238, Paris, France} 

\date{\today}

\begin{abstract}
In the course of evolution, proteins undergo important changes in their amino
acid sequences, while their three-dimensional folded structure and their
biological function remain remarkably conserved. Thanks to modern sequencing
techniques, sequence data accumulate at unprecedented pace. This provides large
sets of so-called homologous, i.e.~evolutionarily related protein sequences, to
which methods of inverse statistical physics can be applied. Using sequence
data as the basis for the inference of Boltzmann distributions from
samples of microscopic configurations or observables, it is possible
to extract information about evolutionary constraints and thus protein function
and structure.  Here we give an overview over some biologically important
questions, and how statistical-mechanics inspired modeling approaches can help to
answer them. Finally, we discuss some open questions, which we expect to be
addressed over the next years. 
\end{abstract}

\maketitle

\section{Introduction}
\label{sec:intro}

Proteins are essential in almost all cellular processes. In the course
of evolution, genomic mutations cause changes in their primary 
amino-acid sequences. However, these changes are not completely
random: only substitutions conserving the biological functionality of a
protein are accepted by natural selection. This leads to an apparently
paradoxical situation: while two proteins of common evolutionary ancestry
(so-called {\it homologs}) may differ in more than 70 or 80\% of their amino
acids, they still have highly similar three-dimensional structure and
biological functions.

Over the last decades, it has been a central issue of biological physics to
relate the amino-acid sequence to its three-dimensional structure, i.e., to
solve the protein-folding problem \cite{bryngelson1995funnels,
  dill2012protein}. To do so, very precise models including the 
physicochemical properties of amino-acids and their detailled interactions have
been designed and simulated \cite{karplus2002molecular}. While
enormous progress has been made in this 
direction (as witnessed by the 2013 Nobel price to Karplus, Levitt and
Warshel), the computational cost of realistic molecular modeling limits the
length of treatable amino-acid sequences and time scales significantly.  

Much more recently, an important alternative based on the progress in
sequencing technology and the increasing availabilty of protein sequences has
emerged.  To date, about 70,000 complete genomes have been sequenced
\cite{mukherjee2017genomes}. Instead 
of considering single amino-acid sequences within a detailled biophysical
model, we can therefore consider entire {\it families of homologous proteins},
i.e. ensembles of diverse sequences believed to have common structure and
function in different species (or different pathways inside the same
species). In current databases many of these families contain more than $10^3$
different sequences and in some cases more than $10^5$ sequences
\cite{finn2016pfam}. {\it The aim 
of the inverse statistical physics of proteins is to capture the sequence
variability in ensembles of homologous sequences, to unveil statistical
constraints acting on this variability, and to relate them to the conserved
biological structure and function of the proteins in this family}.

We start from a multiple-sequence alignment (MSA) of an entire protein
family \cite{durbin1998biological}.
Note that the construction of large MSA is a hard and not completely solved
problem on its own, but here we assume it to be given for the sake of
simplicity. An MSA is a rectangular matrix $A=\{a_i^\mu|i=1,...,L, \mu =
1,...,M\}$, containing $M$ sequences, which are aligned over $L$ positions.
Each entry $a_i^\mu$ of the matrix is either one of the 20 natural amino acids,
or the alignment gap ``-'' introduced to treat amino-acid insertions or
deletions in some sequences. For simplicity, we will consider the gap as a 21st
amino acid throughout this article, and speak about $q=21$ amino acids. Each
row of the MSA $A$ is thus a single protein sequence and each column a specific
position in the proteins (identifiable, e.g., via a specific location in the
conserved three-dimensional protein fold). 

The basic assumption of modeling this MSA using inverse methods
from statistical physics
\cite{roudi2009ising,sessak2009small,mezard2009constraint,cocco2011high,
cocco2011adaptive, aurell2012inverse,nguyen2012bethe,
nguyen2012mean, decelle2014pseudolikelihood} is that it constitutes a
(not necessarily identically and independently distributed) sample of a
Boltzmann distribution (inverse temperature $\beta=1$ without loss of 
generality)
\begin{equation}
\label{eq:boltzmann}
P(a_1,...,a_L) = \frac 1{\cal Z} \exp\{ - \beta {\cal H}(a_1,...,a_L) \}
\end{equation}
associating a probability to each full-length amino-acid sequence
${\bf a}=(a_1,...,a_L)$. While this assumption is a simplification
of the biological reality, it will provide a useful and mathematically
well-defined way to to extract information from data.

The main task of inverse statistical physics is to reconstruct the
Boltzmann distribution $P$ using a sample $A$ drawn from $P$. 
However, in the specific situation of protein sequences this task is
complicated by two opposing facts: On one side, we do not know the 
correct analytical form of the Hamiltonian ${\cal H}$ (e.g., in terms
of local fields, two-spin or higher-order couplings etc.), i.e. a
priori $q^L-1$ free parameters are to be inferred. On the other side,
even the largest MSA cover only a tiny fraction of all possible $q^L$
sequences. Parameter-reduced models have to be used in order to avoid
overfitting. 

We can solve these problems at least partially by assuming a specific
analytical form of ${\cal H}$ and inferring the numerical values of
its defining parameters (e.g., local fields,  pairwise couplings), for
example by maximum 
likelihood. Alternatively, we can decide on a number of observables
(e.g., frequencies, pairwise correlations) whose data-derived
empirical values should coincide with the thermodynamic averages in
our model given by ${\cal H}$, and use the maximum-entropy
principle \cite{jaynes1957information} to fix the analytical form of
${\cal H}$. Both cases contain 
an important subjective element: the decision on a specific
model is largely driven by data availability (more data allow for
more parameters) and the biological question under study. To
illustrate this point, we will list a series of more and more
involved questions, together with an idea about the adequate level
of statistical modeling.  

{\bf 1) Homology detection and sequence annotation} belong to the
classical bioinformatic questions and are usually answered using
statistical models of aligned sequences
\cite{durbin1998biological}. Given a {\it new  natural 
  sequence of unknown biological function}, can we assign it to a
known family of homologous proteins, i.e.~to a given MSA? As discussed
before, these families tend to conserve important parts of the
biological structure and function, so the assignment of the new
sequence to a previously known protein family automatically provides a
prediction of its potential role in biology. Classically homology
detection is done using so-called {\it profile models}, which are
equivalent to {\it independent-site Potts models}  in a heterogeneous
external field,  
\begin{equation}
\label{eq:para}
{\cal H}(a_1,...,a_L) = - \sum_{i=1}^L h_i(a_i)\ .
\end{equation} 
The local fields are able to capture site-specific patterns of {\it
  amino-acid conservation}. This may pertain to a single possible amino
acid in the active site of a protein (corresponding to a column in
the MSA), or to conserved physical amino-acid properties
(e.g., hydrophylic amino acids in an exposed position at the protein's
surface, or hydrophobic amino acids buried inside its core). A more
sophisticated version of these models, called profile Hidden Markov
Model \cite{eddy1998profile}, is able to analyze unaligned sequences
and detect insertions and deletions while maintaining the
independence of amino acids at different aligned positions. 

{\bf 2) Protein-structure prediction and the topology of
  coevolutionary networks}: The assumption of independence in profile
models limits the amount of information that can be
extracted from an MSA since, in practice, amino acids at different
positions do 
not evolve independently. Most single-site mutations are deleterious and
often perturb the physical compatibility with the
surrounding residues in the folded protein. One may imagine that
compensatory mutations in neighboring sites may repair the damage done
by the first mutation; we say that {\it residues in contact coevolve}
\cite{de2013emerging}.

\begin{figure}[h!]
          \centering
          \includegraphics[width=0.4\textwidth]{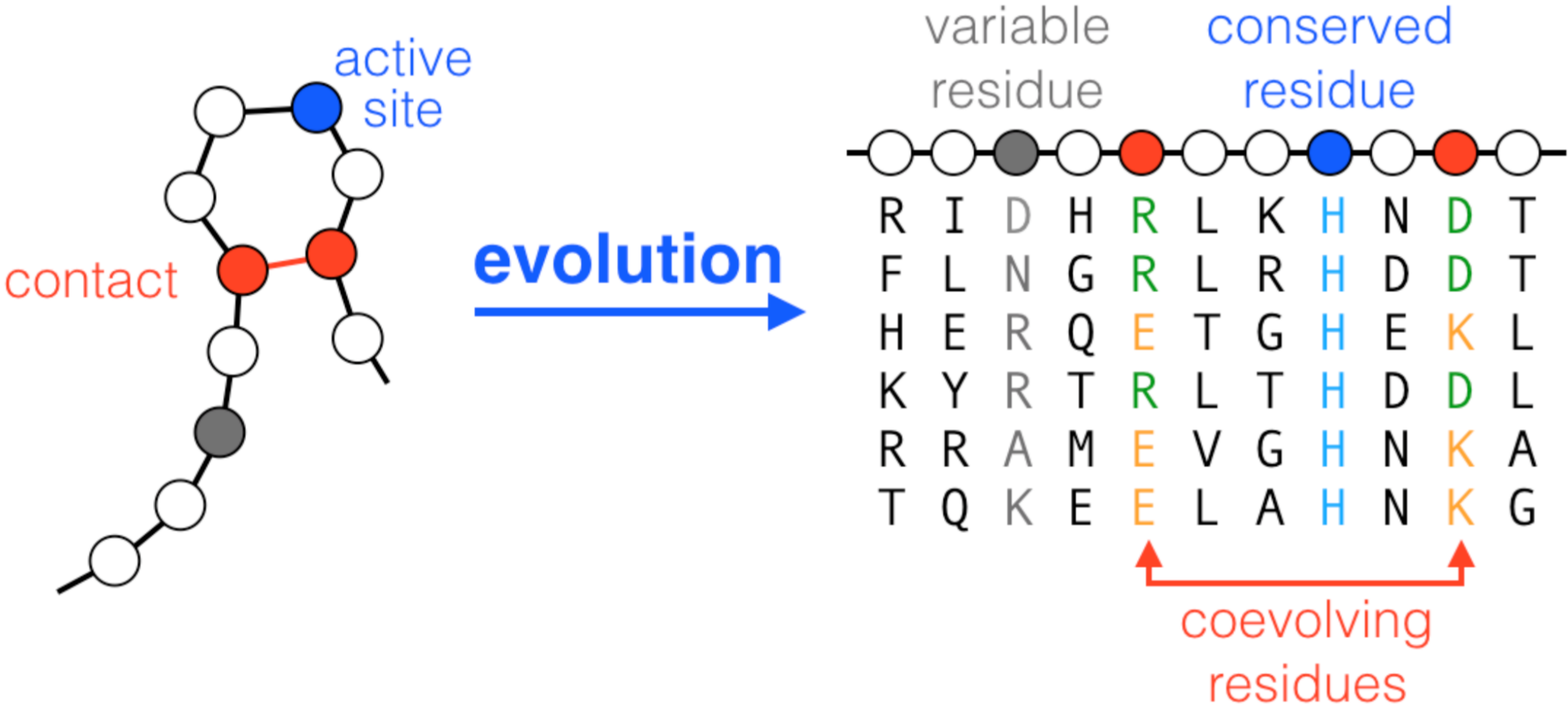}
          \caption{Evolutionary constraints shaping the variability
            between homologous sequences: While constraints on
            individual residues 
            (e.g., active sites) lead to variable levels of amino-acid
            conservation, the conservation of contacts leads to the
            coevolution of structurally neighboring residues and
            therefore to correlations between columns in a
            multiple-sequence alignment of homologous proteins (here
            an artificial alignemt is shown for illustration). }
 \label{fig:coevolution}
  \end{figure}

Coevolution becomes visible in correlated occurrences of amino acids in
different sites, i.e.~via covariation between different columns of the
underlying multiple-sequence alignment
(cf.~Fig.~\ref{fig:coevolution}) \cite{gobel1994correlated,neher1994frequent}. It is
tempting to use such correlations to reconstruct the contact map of a protein,
which could then be used to predict the protein fold as a
three-dimensional embedding of 
this contact map. This idea, present in the literature for more
than 20 years \cite{gobel1994correlated,neher1994frequent}, did
unfortunately not work out easily. A major reason is that correlation (as 
detected in the MSA) is not coupling (as resulting from amino acids in
contact): if, e.g., position $i$ is in contact with position $j$, and position
$j$ in contact with position $k$, we might expect an indirect correlation also
between $i$ and $k$. The aim of the {\it Direct-Coupling Analysis}
\cite{weigt2009identification,morcos2011direct}, and of 
closely related approaches
\cite{burger2010disentangling,balakrishnan2011learning,jones2012psicov},
is to explain correlations via a network of direct 
coevolutionary couplings, or more precisely, via a generalized Potts model
\begin{equation}
\label{eq:potts}
{\cal H}(a_1,...,a_L) = - \sum_{1\leq i<j\leq L} J_{ij} (a_i,a_j) - \sum_{i=1}^L h_i(a_i)
\end{equation} 
containing both local fields $h_i$ and pairwise couplings $J_{ij}$.

As will be shown below, the strongest pairwise couplings provide
accurate predictions of contacts between residues. This enables
protein-structure prediction without the detailed biophysical modeling
mentioned before
\cite{schug2009high,marks2011protein,hopf2012three,nugent2012accurate,
sulkowska2012genomics,ovchinnikov2017protein}. The inference of the
couplings is, however, a 
computationally hard task, since the exact calculation of thermodynamic
averages (which have to coincide with empirical ones) requires a sum
over the exponentially large sequence space (in a disordered model
lacking a priori any symmetry). However, any method reproducing the
{\it topology} of the coevolutionary network underlying
Eq.~(\ref{eq:potts}), i.e. identifying the strongly pairs, is equally
valid for predicting the contact map.

{\bf 3) Inference of mutational landscapes and quantitative sequence
  models}: Quantifying the effect of mutations is a task of
outstanding biomedical importance and can be used as a technique for
identifying causative 
mutations in genetic disease or cancer, or adaptive mutations leading
to therapeutic drug resistance. In a general mathematical setting,
mutational effects can be described by a mutational
landscape, or genotype-phenotype mapping, which associates a
quantitative phenotype $\Phi(a_1,...,a_L)$ to each amino-acid sequence
$(a_1,...,a_L)$ \cite{Visser2014Empirical}. Multiple-protein alignments (of patient derived
sequences or of homologous protein families) have been used to infer
such landscapes from the empirically observed sequence variability
\cite{chakraborty2014hiv,morcos2014coevolutionary,figliuzzi2015coevolutionary,
hopf2017mutation,feinauer2017context},
using in particular the analytical form of the Potts model
Eq.~(\ref{eq:potts}). In this context, the couplings $J_{ij}(a_i,a_j)$
represent so-called {\it epistatic couplings} between mutations.

To quantify the effect of the mutation from amino acid
$a_i$ to $b$ in position $i$, we can calculate the energy
difference
\begin{eqnarray}
\label{eq:deltaE}
\Delta E( a_i\to b) &=& {\cal H}(a_1,...,a_{i-1},b,a_{i+1},...,a_L) 
\\
&&- {\cal H}(a_1,...,a_{i-1},a_i,a_{i+1},...,a_L)
\nonumber
\end{eqnarray}  
between the mutated and the unmutated sequences. Decreasing energies
can be interpretated as potentially beneficial mutations, increasing
energies as potentially deleterious mutations. Contrary to
residue-residue contact predictions inferring the topology of the
coevolutionary network is not sufficient any more. We need 
a {\it quantitative inference} of the local fields and couplings, and expect a -- possibly
non-linear -- correlation between energy differences and mutational 
effects. 

{\bf 4) Protein design and generative sequence models}: In a seminal
work, Ranganathan and coworkers
\cite{russ2005natural,socolich2005evolutionary}  suggested that
the pattern of pairwise residue  
covariation is actually sufficient to generate artificial but fully
functional protein sequences. The basic idea was to shuffle an MSA via
a Monte Carlo procedure such that the statistics of single columns
(residue conservation) and column pairs (coevolution) remains close to
unchanged. In experimental tests, the authors found a substantial
fraction of functional non-natural proteins, whereas imposing only the 
single-column statistics resulted in non-functional amino-acid
sequences. 

In the context of the modeling proposed here, we speculate therefore
that the generalized Potts model Eq.~(\ref{eq:potts}) is 
sufficient to design artificial proteins by Monte Carlo sampling,
while the  profile model Eq.~(\ref{eq:para}) is not. For
this to be true, the inferred model needs to be {\it generative}. This
means that the model must not only reproduce perfectly the empirical
single- and pair-column statistics (within the possibilities of the
fitting algorithm), but should generate sequences that are indistinguishable
from the natural sequences also in higher-order characteristics, which are
not explicitly fitted using Eq.~(\ref{eq:potts}). The
abovementioned results \cite{russ2005natural,socolich2005evolutionary}
suggest that this can 
be actually achieved using models including only local fields and
pairwise couplings, while fields on their own are insufficient.


\section{Statistical Physics of the Inverse Potts Problem}
\label{sec:statmech}

We start from data given as a multiple-sequence alignment
$A=\{a_i^\mu|i=1,...,L, \mu =  1,...,M\}$, which contains $M$ sequences of
aligned length $L$. We model the statistical variability in these data  using
generalized Potts models defined in Eqs.~(\ref{eq:para}) or
(\ref{eq:potts}). In this section we will provide some background on the
underlying inference methodology that connects data and model, in particular on how to
infer the model parameters (local fields $h_i(a)$ and
couplings $J_{ij}(a,b)$).

This methodology will allow us to infer a specific model for each
family of homologous proteins. The corresponding MSA are freely
available in public data bases, such as Pfam \cite{finn2016pfam}.

\subsection{ From Data to Empirical Amino-Acid Frequencies}

The statistical modeling of protein sequences
relies on the fitting of the {\it empirical low-order statistics of the
MSA} $A$, where each row is a protein sequence, and
each column a specific aligned residue position in the proteins. In
particular we estimate the frequencies of the
occurrence of single amino acids $a$ in the MSA column $i$, and of the
co-occurrence of amino acids $a$ and $b$ in columns $i$ and $j$ inside
the same protein:

 \begin{eqnarray}
\label{eq:fi_fij}
 f_i(a)&=&\frac 1{M_{eff}}\sum_{m=1}^M \, w_m \,\delta_{a,a_i^m} \ , \\
 f_{i,j}(a,b)&=& \frac 1{M_{eff}}\sum_{m=1}^M  \, w_m
                 \,\delta_{a,a_i^m}  \,\delta_{b,a_j^m} \nonumber  \ , 
\end{eqnarray}
with $\delta_{a,b}=1$ being the Kronecker symbol, which equals
one if amino acids $a$ and $b$ are equal, and zero otherwise. 

Note that in the sum we have introduced a sequence-specific weight
$w_m$, as well as the {\it effective sequence number} $M_{eff} =
\sum_mw_m$ \cite{morcos2011direct}. 
The reason is that sequences in an MSA {\it cannot be considered as
independent} configurations drawn from the statistical model $P$.  The MSA
collects homologous sequences that have a common ancestry. This ancestry is
often relatively recent, and as a consequence sequences can be atypically
similar to each other because there was no time for them to evolve further
apart. While this similarity can be used to reconstruct the evolutionary
history (i.e.~the phylogenetic tree) of these sequences
\cite{felsenstein2004inferring}, in terms of 
probabilistic modeling it is a sampling bias introducing correlations not
related to the structure and function of the proteins. A further bias
results from the uneven selection of sequenced species -- species close to
model species or important pathogens tend to be more frequently sequenced than
species without any direct scientific or biomedical interest.

While a proper removal of this bias remains an important open
question, it can be counterbalanced by reducing the weight $w_m$ of
similar sequences in the empirical frequency counts in
Eq.~(\ref{eq:fi_fij}). In a commonly used definition of the weights in 
the context of residue-contact prediction,  $w_m$ equals to the
inverse of the number of sequences ${\bf a}^\ell$  of Hamming distance
$d_H({\bf a}^\ell, {\bf a}^m)$ (number of positions with distinct amino
acids) smaller than $xL$.  It has been shown that $x\simeq 0.2-0.3$ leads to
accurate and robust results. Note that $\ell = m$ has to be counted,
too, so the weight for an isolated sequence is $w_m=1$, while it is
smaller for any close-to-repeated sequence \cite{morcos2011direct}. 

The empirical frequencies of Eqs.~(\ref{eq:fi_fij}) allow
for quantifying the level of correlation between the amino-acid
occupancies of any pair of positions $(i,j)$. The mutual
information 
\begin{equation}
\label{eq:MI}
MI_{ij} = \sum_{a,b} f_{ij}(a,b) \log \frac{f_{ij}(a,b)}{f_{i}(a)
  f_{j}(b)}\ .
\end{equation}
is zero if and only if sites $i$ and $j$ are statistically
independent, and positive else. It might be tempting to use the mutual 
information as a simple proxy for pairwise couplings. However, in the
context of proteins this leads to inaccurate results since correlations
emerge from networks of couplings. Inverse methods are
needed to disentangle direct couplings from indirect correlations.

\subsection{From Amino-Acid Frequencies to Potts models} 

After extracting the empirical statistics from the MSA we can
use it to estimate the parameters of the Potts model.
This means enforcing the model to reproduce the
relevant empirical statistics, which can be only the first-order statistics given by
the $f_i(a)$ in the independent-site case or the second-order statistics
given by the $f_{ij}(a,b)$ in the case of a pairwise Potts model. 
The latter constraints lead to a model with pairwise couplings $J_{ij}(a,b)$.
This is expected to be more realistic, but comes at a high computational
cost since exact inference requires the calculation of thermodynamic averages and
thus the partition function ${\cal Z}$ as a sum over an exponential
number $q^L$ of amino-acid sequences ${\bf a} = (a_1,...,a_L)$. In
this subsection we will ignore this technical difficulty, discussing
the general setting of inverse Potts models. We leave the question of
efficient approximations to the next subsection.

\subsubsection{The independent-site case (IND)}

In the profile case of Eq.~(\ref{eq:para}), all positions in the model distribution $P$ are
independent from each other and the joint probability is factorized over
positions. We can therefore easily fit the fields $h_i(a)$ to satisfy
\begin{equation}
\label{eq:para_inf}
f_i(a) = \frac{e^{h_i(a)}}{\sum_b e^{h_i(b)}}
\end{equation}
for all positions $i=1,...,L$ and all amino acids $a$. This equation
can be, up to a constant, easily inverted through $h_i(a) = \ln f_i(a) +
const$. To avoid fields to become (minus) infinite in the case of
unobserved amino acids, a {\it regularization} or a {\it pseudocount}
can be introduced, see below. This model will not
reproduce the second-order statistics $f_{ij}(a,b)$ and the
probability of co-occurrence of amino acids $a$ and $b$ in the
positions $i$ and $j$ is equal to $f_i(a)f_j(b)$ in the model
distribution. 

\subsubsection{ Inference of pairwise Potts models}
The parameters of the pairwise model Eq.~(\ref{eq:potts}), the
local fields $h_i(a)$ and the couplings $\{J_{i,j}(a,b)\}$, have to be
inferred in such a way that first {\it and} second moments of the  
Boltzmann distribution $P$ correspond to the empirical frequencies.  The
following equations have to be satisfied,
\begin{eqnarray}
\label{eq:invmatch}
f_i(a) &=& \langle \delta_{a,a_i} \rangle_P \\
f_{i,j}(a,b) &=& \langle \delta_{a_i,a} \delta_{a_j,b}\rangle_P 
\nonumber 
\end{eqnarray}  
for all $i,j,a,b$. This ensures that the marginal single- and two-site
frequencies of the Potts model are equal to the empirical
frequency counts derived from the original data set.
In this equation,  $\langle \cdot\rangle_P$ denotes the average
over the Boltzmann distribution $P$ in Eq.~(\ref{eq:boltzmann}): 
\begin{equation}
 \langle {\cal O}({\bf a})\rangle_P\equiv  \sum_{\bf a} \,P({\bf
   a})\, {\cal O}({\bf a})
\end{equation}
for any observable ${\cal O}({\bf a})$ depending on the amino-acid
sequence ${\bf a}=(a_1,...,a_L)$. The $N_p = Lq + \frac{L(L-1)}2 q^2$
equations in Eqs.~(\ref{eq:invmatch}) are coupled and have to be solved
simultaneously. 

\subsubsection{Overparametrization and gauge invariance} 
Note that Eqs.~(\ref{eq:invmatch}) are not independent. As
each site $i$ carries one amino acid $a$, the frequencies $f_i (a)$
sum up to one. Hence there are only $L(q-1)$ independent one-point
frequencies. Similarly, for each one of the $\frac{L(L-1)}2$ pairs
$i,j$, only $(q-1)^2$ pairwise frequencies $f_{ij}(a,b)$ are
independent, as marginalizing over $b$ allows us to recover the
single-site frequencies:  $\sum_b f_{ij} (a,b)=f_i(a)$.   
The number of independent constraints coming from
Eqs.~(\ref{eq:invmatch}) is therefore only $N_c=L(q-1) +
\frac{L(L-1)}{2} (q-1)^2$.   

The number of parameters, $h_i(a)$ and $J_{ij}(a,b)$, defining
Hamiltonian ${\cal H}$ in Eq.~(\ref{eq:potts}) equals $N_p$ and,
therefore, exceeds the number of independent equations in
Eq. ~(\ref{eq:invmatch}). This {\it overparametrization} gives rise to
the following {\it gauge invariance}: The probability $P({\bf a})$
remains unchanged under the (gauge) transformation
\begin{eqnarray}
J_{ij}(a,b) &\leftarrow& J_{ij}(a,b) +K_{ij} (a) + K_{ji} (b) \  , \\
h_i(a) &\leftarrow& h_i(a) +g_i -\sum_{j(\neq i)} \big( K_{ij} (a) +
                 K_{ji} (a) \big) \ .  \nonumber  
\end{eqnarray}
Here, $K_{ij}(a)$ and $g_i$ are arbitrary quantities. Note that 
$K_{ij}(a)$ and $K_{ji}(b)$ are not independent since any
$(ij)$-dependent quantity added to the first and subtracted from the
second leaves the transformation invariant. The total number of
independent parameters defining the above transformation equals
therefore $N_q=L+\frac{L(L-1)}2\times (2 q-1) $. 

Since each gauge transformation leaves the Boltzmann distribution $P$
unchanged, the number of parameters to be fitted to the empirical data
actually reduces to $N_p-N_q$. This equals the number $N_c$ of
independent equations in Eqs.~(\ref{eq:invmatch}), and the inference
problem is well defined once the gauge is fixed. Two widely used gauges are:
\begin{itemize}
\item The {\it lattice-gas gauge}, in which $h_i(q)=J_{ij}(a,q) =
  J_{ij}(q,a)=0$ for all $i,j,a$, measures all energies with respect to
  the ``empty'' configuration $(q,...,q)$, and considers $q-1$
  different ``particle'' types $a=1,...,q-1$. 
\item The {\it zero-sum gauge} (or Ising gauge) assumes that $\sum_a
  h_i(a) = \sum_a J_{ij}(a,b) = 0$ for all $i,j,b$. This
  gauge generalizes the well-known case of Ising models
  ($J_{ij}(s_i,s_j) = J_{ij} s_i s_j, h_i(s_i) = h_is_i, s_{i,j}=\pm
  1$) to $q$-state Potts spins. Contrary to the lattice-gas gauge
  above, this gauge does not arbitrarily break the symmetry between the
  $q$ states of the Potts variables. 
\end{itemize}
As is the case with standard Ising and lattice-gas models, the two
formulations are  related through a simple (gauge)
transformation of the underlying local variables, and are therefore equivalent.

\subsubsection{ Cross-entropy minimization and Bayesian interpretation}
In this section we rewrite the inverse problem in Eqs.~(\ref{eq:invmatch})
in a simpler and more interpretable way.  
We introduce the cross entropy 
\begin{eqnarray}
  \label{eq:Sc}
  S_c(\{ h, J \}) & =& - \frac 1{M_{eff}} \sum _m w_m \, \log P(a_1^m,a_2^m,...,a_L^m) 
 \nonumber\\ 
  &=&\log {\cal Z}- \sum_{i,a} h_i(a)\;f_i(a) \nonumber \\
  && -\sum_{i<j,a,b} J_{ij}(a,b)\,f_{ij}(a,b) \ .
  \end{eqnarray}
This quantity gives -- up to an additive $P$-independent constant
equal to the entropy of the empirical distribution
$f(a_1,...,a_L)=\sum_m w_m \prod_i \delta_{a_i,a_i^m}$ of the
sequences in the MSA -- the Kullback-Leibler (KL) divergence
\begin{equation}
\label{eq:dkl}
D_{KL}(f||P) = \sum_{\bf a} f({\bf a}) \log \frac{f({\bf a})}{P({\bf a})} 
\end{equation} 
between the empirical distribution $f$ (of which $f_i(a)$ and
$f_{ij}(a,b)$ are marginals) and the Boltzmann 
distribution $P$. Within the class of pairwise Potts models defined by 
Eq.~(\ref{eq:potts}), a possible parameter choice is the one minimizing
$D_{KL}$ and, hence, the cross entropy $S_c$. It is easy to check that
the minimization with respect to fields $h_i(a)$ and couplings
$J_{ij}(a,b)$ leads back to Eqs.~(\ref{eq:invmatch}). Note that the
Hessian matrix of $S_c$ is nonnegative. Zero eigenvalues correspond
to directions in which the  cross entropy has a minimum for parameters
going to (minus) infinity; they make regularization necessary as
explained in the next section. 
  
After minimization, the cross entropy $S_c$ becomes a function of the
empirical one- and two-point amino-acid frequencies; more precisely it
is the Legendre transform of the negative free energy $\log {\cal
  Z}$. The Potts parameters (fields, couplings) act as conjugated
parameters to observables (one- and two-point frequencies), in the
same way as pressure or chemical potential are conjugated to 
volume and number of particles in thermodynamical 
statistical ensembles.  One of the basic properties of Legendre
transforms is that the derivative of the potential with respect to its
control parameters leads to the conjugate parameters.   

Minimizing of the cross entropy is equivalent to maximizing
the log-likelihood in a Bayesian framework. Bayes' theorem
in fact allows one to express the posterior probability of the model
parameters given the data, $ P(\{h,J\}| \{{\bf a^{m}}\})$, starting
from the probability of the data given the model, $P(\{{\bf a^{m}}\}|
\{h,J\} )$: 
  \begin{equation}
  P(\{h,J\}| \{{\bf a^{m}}\})=\frac { P(\{{\bf a^{m}}\}| \{h,J\}
    )\;P_0\left(\{h,J\}\right)}{P(\{{\bf a^{m}}\})} \ , 
  \label{eq:bayes}
  \end{equation}
where $P_0$ is a prior distribution over the Potts model
parameters. When the prior is uniform and the sequences are
independently drawn from $P$, maximizing the posterior
distribution over all couplings and fields is equivalent to
minimizing the cross entropy $S_c$ with uniform weights
$w_m=\frac 1M$. In the presence of a non-uniform prior a term $-\log 
P_0 (\{h,J\})$ is added to the cross entropy, which plays the role of
a regularization for the Potts parameters. In the example of a Gaussian prior
$P_0$, this penalty term is equivalent to a $L_2$ regularization,
cf. next subsection.

\subsubsection{Regularization} 

Typical proteins (or protein domains, which constitute the alignable modules
making a protein) have a length of $L\simeq 50-500$. This leads to $N_c\simeq
5\cdot 10^5 - 5\cdot 10^7$ parameters to be inferred. Combined with limited
sampling ($M\simeq 10^2-10^5$ for typical MSA), this large number of parameters
makes regularization to avoid overfitting necessary.

Consider as an illustration the case of two amino acids, $a$ and $b$, that are
rarely encountered on their respective sites $i$ and $j$. Assuming their
frequencies to be $f_i(a)=f_j(b)=0.01$ (note that the average frequency of
amino acids is $1/q \simeq 0.048$) and that they evolve independently from each
other, the probability of finding both amino acids in a sequence is equal to
$0.0001$. For MSAs with less than 10,000 sequences, such a sequence is
typically not encountered. This results in an apparent anticorrelation, and
ultimately, an infinitely negative coupling between the two sites and amino
acids. On the contrary, if the combination is found in a single out of much
less than 10,000 sequences, it will lead to a large positive but statistically
unsupported coupling. 

To avoid such sampling effects, various
regularization schemes can be used. In practice, a
{\it penalty} $\Delta S_c(\{h,J\})$ is added to the cross entropy
(\ref{eq:Sc}) during the minimization of the parameters
$\{h,J\}$. Standard examples are the $L_1$ norm, where  
  \begin{equation}
  \Delta S_c^{L_1}(\{ h,J \})=\gamma_1 \sum_{i,a} |h_i(a)|+ \gamma_2
  \sum_{i<j,a,b} |J_{ij}(a,b)| \ .
  \end{equation}
The non-analyticity in zero penalizes small fields and couplings,
forcing them to become exactly zero in value and thus 
favors sparse networks to be inferred. The $L_2$ norm, 
  \begin{equation}
  \Delta S_c^{L_2}(\{ h,J \})= \gamma_1 \sum_{i,a} h_i(a)^2+ \gamma_2
  \sum_{i<j,a,b} J_{ij}(a,b)^2 \ ,
  \label{l2}
  \end{equation}
penalizes large absolute values of parameters. This is found to be
efficient in the context of protein MSA since it removes the
spuriously large parameter values based on insufficient sampling.

Another frequently used procedure to limit undersampling effects
consists of adding a {\it pseudocount} to the empirical one- and
two-point frequencies (justifiable via a Dirichlet prior on
frequency counts):  
\begin{eqnarray}
 f_{i}(a) &\leftarrow& (1- \alpha) \, f_{i}(a) +\frac{\alpha}{q} \ , \nonumber \\
 f_{ij}(a,b) &\leftarrow&  (1- \alpha)\, f_{ij}(a,b) +\frac{\alpha}{q^2} .
 \end{eqnarray}
The introduction of a pseudocount is -- up to finite-sample effects --
equivalent to assuming that the MSA 
is extended with a fraction $\alpha/(1-\alpha)$ of sequences with
amino acids sampled uniformly. Regularization parameters
$\gamma_1,\gamma_2 $ and $\alpha$ should in principal vanish as the
number of data increases; values for these parameters used in practice
will be discussed later on.  An alternative way to regularize the inference
problem is to reduce the number of Potts states and keep well sampled
amino acids only, for example with frequencies larger than some
threshold values. The number of Potts states then depends on the
site. This can drastically reduce the number of parameters to infer,
and a weaker regularization on the 
remaining parameters can be used. 
  
\subsection{Methods of Approximate Inference} 
The convexity of the properly regularized cross entropy guarantees the
uniqueness of the minimum and the optimal parameter values ${h,J}$. This minimum
can be found by local convex optimization techniques. A problem is that
calculating the averages in Eqs.~(\ref{eq:invmatch}) exactly requires the 
calculation of the partition function ${\cal Z}$. This is
in practice intractable for proteins with hundreds of residues since it includes a
sum over all $q^N$ sequences. We therefore resort to
approximation schemes to solve the inference problem 
\cite{roudi2009ising,sessak2009small,mezard2009constraint,cocco2011high,
cocco2011adaptive, aurell2012inverse,nguyen2012bethe,
nguyen2012mean, decelle2014pseudolikelihood}. Their relative
advantages and disadvantages in the context of protein sequences will
be discussed in the {\it Results} section.
  
\subsubsection{Boltzmann Machine Learning (BM)} 
The most straightforward approximation to solve
Eqs.~(\ref{eq:invmatch}) is called {\it Boltzmann machine
  learning} \cite{ackley1985learning}. Starting from an initial guess
for the values of the couplings  
and fields (e.g., the solution of the independent-site model described
above), the one- and two-point marginals of $P$ are estimated by Monte
Carlo Markov Chain (MCMC) sampling. They are then compared with the
empirical one- and two-point frequencies, and the fields and couplings
are modified according to  
   \begin{eqnarray}
\label{eq:inversebm}
 h_i(a) &\leftarrow & h_i(a) + \epsilon \big( f_{i}(a)-\langle \delta
                      _{a_i,a}\rangle_P \big) \ , \\ 
J_{ij}(a,b) &\leftarrow& J_{ij}(a,b) + \epsilon \big(
                          f_{ij}(a,b)-\langle \delta _{a_i,a}\,
                          \delta_{a_j,b}\rangle_{P}\big) \nonumber \ , 
\end{eqnarray}  
where $\epsilon$ is a small parameter. The direction of the update follows the
gradient of the cross entropy $S_c$. In the case of sufficiently precise MCMC
sampling and a sufficiently small $\epsilon$, this iterative procedure is
guaranteed to converge towards the solution of the fixed-point
equations~(\ref{eq:invmatch}).
 
This procedure cannot be used directly due to its high
computational cost. Efficient implementation going beyond simple
gradient descent makes Boltzmann machine learning applicable to protein
families smaller than about $L=200$
\cite{sutto2015residue,haldane2016structural,barrat2016improving}. Large-scale
studies for hundreds or 
even thousands of protein families remains therefore out of reach.

\subsubsection{ Gaussian Approximation}  
Using the lattice-gas gauge mentioned before, we can represent each
Potts variable $a$ by a $(q-1)$-dimensional vector
$(s^1,...,s^{q-1})$ with entries $s^b(a) = \delta_{b,a}$. Amino acids
$a=1,...,q-1$ are thus represented by unit vectors in direction $a$,
while the reference amino acid (or the gap) is represented by the zero
vetor $(0,...,0)$. The MSA becomes a matrix of $M$ rows and $(q-1)L$
columns with binary entries $\{0,1\}$. The average of the column
corresponding to position $i$, amino acid $a$ equals the empirical
frequency $f_i(a)$, the $(q-1)L$-dimensional covariance matrix 
$C_{ij}(a,b)=f_{ij}(a,b)-f_i(a)f_j(b), i,j=1,...,L, a,b=1,...,(q-1)$. 

The Gaussian approximation ignores the binary nature of the
$s$-variables and treats them as continuous variables having the
same means and covariances \cite{jones2012psicov,baldassi2014fast}. 
The pairwise Potts model is transferred into a multivariate Gaussian
model with parameters $J_{ij}(a,b)$. The cross-entropy can be
calculated analytically (up to a coupling-independent
normalization),
\begin{equation}
 S_c^G=-\frac{1}{2} \log \text{Det}{(- J)}-{\sum_{i<j,a,b}
 } J_{ij}(a,b)\;C_{ij}(a,b)\ .
 \label{eq:gausss}
\end{equation}
Eq.~(\ref{eq:gausss}) can be easily minimized  over $J$, giving
\begin{equation}
\label{eq:Cinv}
  J_{ij}(a,b)=-\left(C^{-1}\right)_{ij}(a,b) \ .
\end{equation}
The original exponential-time inference problem (time $~q^L$) is
replaced by a simple inversion of the empirical covariance matrix, a
task requiring $\sim L^3$ operations. It can be
achieved on a standard desktop computer even for long proteins of
$L\simeq 1000$ amino acids.

\subsubsection{Mean-field Approximation (MF)}
The standard MF approximation allows one  to estimate
one-point marginals self-consistently. Plugging in
Eqs.~(\ref{eq:invmatch}), which equate empirical and model derived
frequencies, we find 
\begin{equation}
\label{eq:mf}
\frac{f_i(a)}{f_i(q)} = \exp\left\{h_i(a) + \sum_{j,b} J_{ij}(a,b)
  f_j(b)\right\} 
\end{equation}
within the lattice-gas gauge. Covariances can be calculated using linear
response, i.e., $\partial P_i(a) / \partial h_j(b) = C_{ij}(a,b)$. This again
leads to Eq.~(\ref{eq:Cinv}) and the couplings can be obtained by inverting the
covariance matrix \cite{roudi2009ising,morcos2011direct}. The fields can then be
obtained by resolving Eqs.~(\ref{eq:mf}) with given single-site
frequencies and couplings. 

Together with the Gaussian approximation, MF is currently the computationally
most efficient approximative inference scheme for the inverse Potts problem.
However, it will not converge to the exact solution even with infinite data, and
no rigorous bounds on inference errors are known. One can also formulate
refined mean-field approximations, based on e.g., the Thouless-Anderson-Palmer
or Bethe-Peierls approximations, but they have not found to be of advantage
when applied to protein MSAs.

\subsubsection{Pseudolikelihood Maximization (PLM)} 

In the PLM technique
\cite{ravikumar2010high,aurell2012inverse,ekeberg2013improved}, the 
log probability of the data given the parameters in 
Eq.~(\ref{eq:Sc}) is replaced by a sum of site-dependent terms:

\begin{equation}
\sum\limits_{m=1}^{M} w_m \log P(a_1^m,...,a_L^m) \rightarrow
\sum\limits_{i=1}^{L} \sum\limits_{m=1}^{M} w_m \log P(a_i^m | {\bf
  a}_{-i}^m) 
\end{equation}

Here, $i$ denotes a site and ${\bf a}_{-i}$ denotes the sequence $a$ without the $i$th
site. Note that the sum contains one term for each $i$ and these terms contain
probability distributions over a single amino acid (given the others).  This
means that for normalizing these terms we need to calculate $L$ individual sums
over $q$ amino acids instead of one sum over $q^L$ sequences. Setting 

\begin{equation}
S_c^{PLM}(i) = - \sum\limits_{m=1}^{M} w_m \log P(a_i^m | {\bf a}_{-i}^m),
\end{equation}
the cross entropy is the sum

\begin{equation}
S_c^{PLM} = \sum_{i=1}^L S_c^{PLM}(i)
\end{equation}
over the site-dependent terms. In terms of couplings and fields these terms can be written as
\begin{eqnarray}
S_c^{PLM} (i)&=& \sum_{m=1}^{M} w_m\left[ \log\left\{\sum_a
                 e^{h_{i}(a)  +\sum_{j(\neq i)} J_{ij} (a,a_j^{m}) }
                 \right\} \right.\nonumber \\ 
&& 
-  h_{i} (a_i^{m}) - \left. \sum_{j(\neq i)}  J_{ij} ( a_i^{m}, a_j^{m}) \right]\ .
\end{eqnarray}

While $S_c^{PLM}$ is not an approximation for the cross-entropy $S_c$, the
method of minimizing $S_c^{PLM}$ can be shown to be {\it statistically
consistent} \cite{ravikumar2010high}: The true couplings and fields
are recovered in the limit of 
infinite data if the data are indeed sampled from a pairwise Boltzmann
distribution. 

Interestingly, this statistical consistency remains valid if the $S_c^{PLM}(i)$ are minimized
independently. While this allows for a parallel and efficient implementation,
it leads for finite data to asymmetric couplings $J_{ij}(a,b) \neq
J_{ji}(b,a)$. In practice, we infer each coupling as the mean of the asymmetric estimates
$[J_{ij}(a,b) + J_{ji}(b,a)]/2$.

The key advantage of PLM is
that the cross entropy is calculated from the sampled sequences, and there is
therefore no need for an exponential-time calculation of the partition
function.  However, $S_c^{PLM}$ does not depend only on the empirical
single- and two-site frequencies, but on the complete set of sampled
configurations. PLM becomes therefore slower for larger samples and the running
time grows linearly with the number $M$ of sequences. 

\subsubsection{Adaptive Cluster Expansion (ACE)}
The minimal cross entropy $S_c$ can be formally written as a sum of
$2^L-1$  cluster contributions \cite{cocco2011adaptive}, each
depending only on  the  empirical 
frequencies $f_i$ and $f_{ij}$  of the sites inside the cluster: 
\begin{eqnarray}
\label{eq:ace}
S_c^{ACE} &=&\sum_{i} \Delta S_i (f_i) + \sum_{i<j} \Delta S_{ij} (f_i,f_j,f_{ij})  \\
&+& \sum_{i<j<k} \Delta S_{ijk} (f_i,f_j,f_k,f_{ij},f_{jk}, f_{ik}) + \ldots\nonumber
\end{eqnarray}
Here, $\Delta S$ denotes the contribution to the cross entropy from a
cluster that cannot be deduced from the contributions of all its
subclusters \cite{cocco2012adaptive}. Expansion (\ref{eq:ace}) is, however, of limited
use if not truncated to a small number (compared to $2^L$) of
terms.  A convenient truncation scheme is obtained by summing only
contributions $\Delta S$ exceeding (in absolute value) an arbitrary
threshold $\theta$, as large cluster contributions automatically
identify groups of strongly interacting variables. In practice,
clusters are recursively constructed, starting from small size ones (2
sites) through progressive inclusion of sites. The process is iterated
until all newly created clusters have $|\Delta S|$ smaller than
$\theta$. The derivatives of $S_c^{ACE}$ with respect to the frequencies
$\{f_i(a),f_{ij}(a,b)\}$ give access to the Potts parameters
$\{h_i(a),J_{ij}(a,b)\}$. The threshold $\theta$ is tuned such that
the inferred model reproduces the data statistics,  as can be verified
through Monte Carlo sampling, up to finite-sampling errors.  

ACE, similarly to Boltzmann-machine learning and differently from the
Gaussian, mean-field and pseudolikelihood approximations, accurately
reproduces the  sampled frequencies and correlations by construction
\cite{cocco2012adaptive,barton2016}. 
Due to the properties of the inverse susceptibility (Fisher information) 
matrix, the convergence of the expansion depends on the structure of 
the interaction graph rather than on the magnitude of data
correlation. The threshold $\theta$ acts as a regularizer:  the
complexity of the network is adapted to the level of sampling, in
particular it will be sparse for bad sampling. The disadvantage of ACE
is that it is computationally more involved than any of the other
approximative methods: it requires the exact calculation of
the partition function for each cluster.  The computational burden is
largely decreased by the compression of the number of Potts states to
restrict the inferred model to only well observed amino acids, as
described above, and by the possibility to run it in combination with
Boltzmann-machine learning \cite{barton2016}. 
   
\subsubsection{Other Approaches} 
There are several other interesting methods, {\em  e.g.} minimal
probability flow \cite{mpf2011} or Bayesian networks
\cite{burger2010disentangling}, which have been applied 
to protein sequence modelling. 

\section{Modeling Families of Protein Sequences}

After having adressed inverse models in statistical physics from the
methodological point of view, we will give a few exemplary results from using
these methodologies with real protein sequence data.

\subsection{Homologous Protein Families and Profile Models}

State-of-the-art techniques for homology detection and sequence alignments
are based on the patterns of amino-acid conservation in individual
positions \cite{durbin1998biological}. A
class of statistical models are independent-site {\it profile models}, which assume
all positions (or columns in an alignment) to be statistically independent. In
the case of unaligned sequences, profile models can be generalized to {\it
profile hidden Markov models}. These combine residue conservation patterns with
the possibility of identifying insertions and deletions in a sequence and
are among the most successful statistical models in bioinformatical
sequence analysis. They are at the basis of databases of homologous protein
families such as Pfam. The Pfam database (release 28.0 in 2015) contains 16,230 protein domain
families, out of which 6,783 contain more than 1,000 sequences
\cite{finn2016pfam}. This number of 
sequences is empirically known to be a lower bound to the size of an input MSA
providing accurate and robust predictions by pairwise statistical modeling.
Only 10 years ago, there were about 600 families (release 20.0 in 2006)
in Pfam, and the number of accessible families continues to grow quickly.
Approaches from inverse statistical physics that go beyond profile
models can now be applied to the majority of protein families. 

\subsection{Potts models: accurate fitting vs. topological inference} 

\begin{figure}[h!]
   \centering
   \includegraphics[width=0.5\textwidth]{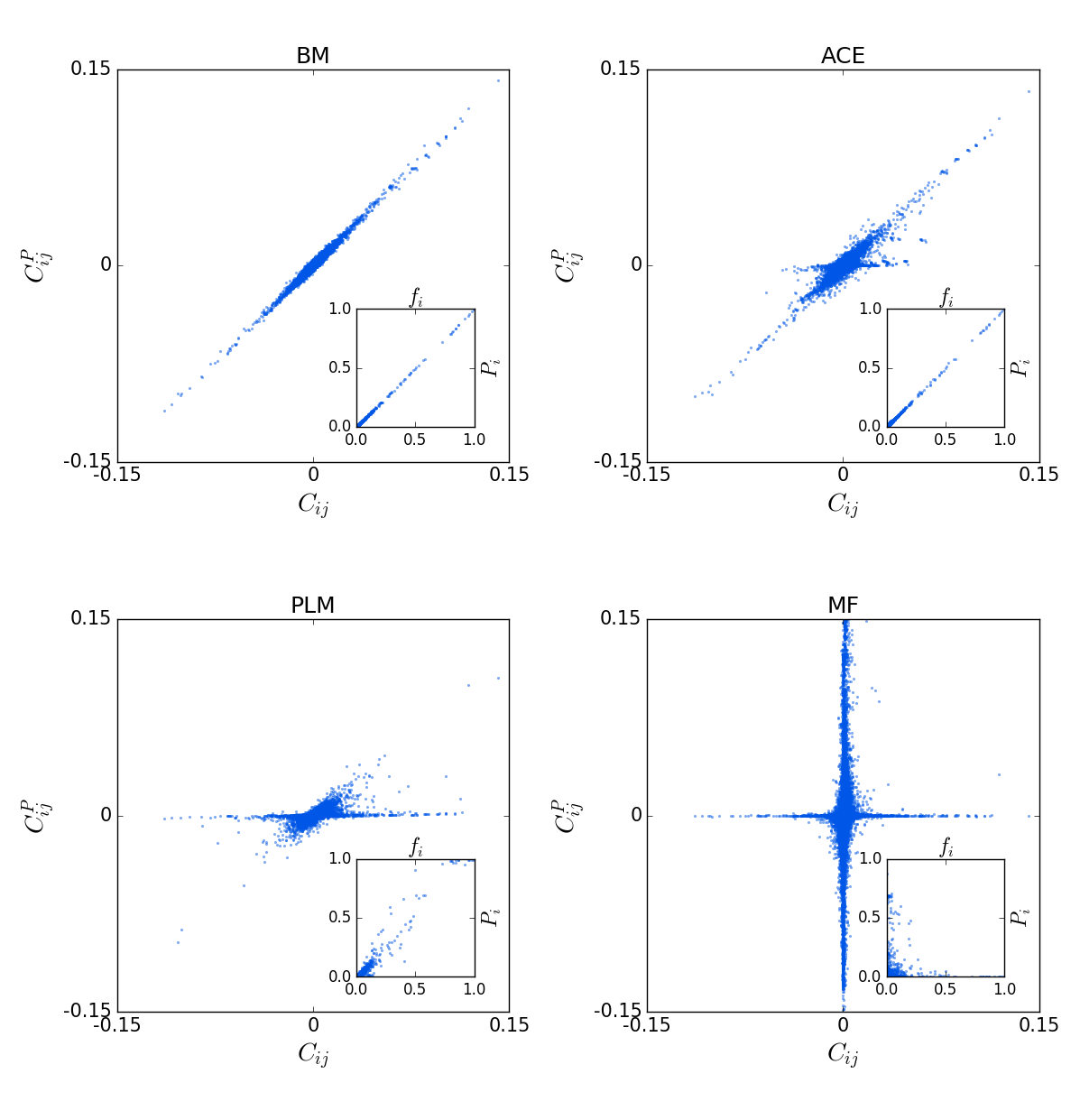}
   \caption{Comparison of different methods: Covariances (main panels)
     and single-point frequencies of the model P as commpared to the
     empirical values resulting from the original MSA. While the
     Boltzmann machine (BM) is guaranteed to accurately fit the
     empirical values after e sufficient number of iterations of
     Eq.~(\ref{eq:inversebm}), all others are based on
     approximations. MF does not reproduce the empirical statistics
     (it actually has important ergodicity problems when sampling by
     MCMC), while PLM shows a clear improvement. ACE accurately fits
     the single-sire frequencies and large enough covariances,
     deviations for small covariances result from lumping together
     all symbols $a$ with probabilities $f_i(a)<0.05$,   such covariances have therefore  low
     statistical support and large relative error,
     cf.~\cite{barton2016}).}
\label{fig:fi_fij_fapc}
\end{figure}

The inference of the couplings and fields of a pairwise Potts model
(\ref{eq:potts}) from aligned amino-acid sequences is a computationally hard
task. It is expected that the approximate inference methods vary in quality of the resulting predictions. Since a ``true'' model
does not exist -- evolution is not a Monte Carlo simulation of a Boltzmann
distribution -- it is not clear how to assess the accuracy of the inferred
model. A straight forward benchmark would be to check if
Eqs.~(\ref{eq:invmatch}), which equate empirical and model statistics, are well
fitted. Using the protein family PF00014 (Trypsin inhibitor: $L=53, M=4915$) as
an example, we apply the different approaches discussed in the last section for
parameter inference. Subsequently we sample artificial sequences from the
inferred statistical model $P(a_1,...,a_L)$ using MCMC, in order to estimate the
model statistics. In Fig.~\ref{fig:fi_fij_fapc} we show the results: While mean-field
inference \cite{morcos2011direct} and pseudo-likelihood maximisation
\cite{ekeberg2013improved} reproduce the empirical
statistics badly (mean-field actually leads to serious problems in equilibration MCMC
simulations), BM \cite{figliuzzi2017} and ACE \cite{barton2016ace} are
much more precise. 

Does this mean that MF and PLM are of no use for analysing large MSA? As argued
in the introduction, the answer is not so easy. When predicting residue
contacts in the three-dimensional fold of a protein, we do not need the fine
statistics of the empirical data to be reproduced. We need to correctly capture
the {\it topology of the network of coevolutionary couplings}. 

To assess the topology, we need to map the $q \times q$ coupling matrix
$J_{ij}(a,b)$ for each pair $(i,j)$ onto a scalar quantity measuring
the {\it coupling strength} between the two sites $i$ and
$j$. 
A quantity often used \cite{ekeberg2013improved} is the {\it Frobenius
  norm} of the coupling matrices for each $(i,j)$,   
\begin{equation}
F_{ij}=\sqrt{\sum_{a,b} J_{ij} (a,b)^2} \ .
\end{equation}
Since the Frobenius Norm is gauge dependent, we have to specify in which gauge
we calculate it. A sensible choice is the zero-sum gauge discussed above. It
minimises the Frobenius norm and explains thereby ``as much as possible'' by
fields, and ``as little as necessary'' by pairwise couplings. Contact
predictions improve furthermore when using the {\it average-product correction}
(APC) \cite{dunn2008mutual}: 
\begin{equation}
\label{eq:apc}
F_{ij}^{APC} = F _{ij}- \frac{\sum_k F_{ik}  \sum_k F_{kj}} {\sum_{k,\ell} F_{k\ell}} \ .
\end{equation}
This correction is identical to the measure of modularity commonly
used for describing networks \cite{Hugo2015}, and amounts to substracting
from the Frobenius norm a null-model contribution for the pair
$i,j$ due to the single-site properties of $i$ and $j$ \cite{Newman2006}. 

The resulting $F^{APC}$-values are highly correlated for different
methods. In particular the largest values, which characterize the
sub-network of the strongest coevolutionary couplings, are found to
strongly overlap. As an example, out of the top 50 coupled pairs found
by BM, 80\% (resp. 70\% / 62\%) are also within the first 50 pairs
identified by PLM (resp. MF / ACE); these fractions rise to 100\%
(resp. 88\% / 76\%) if we look for these 50 top-scoring BM predictions within
the first 100 PLM (resp. MF / ACE) pairings. The lower overlap between ACE and
the other methods results from the weaker regularization used, which
allows for a better fitting but at the cost of introducing a number of
strong couplings between rare amino acids. A random selection of 50
pairs for each method would result in an mean overlap of less than
4\%, illustrating the high significance of the reported overlaps.


We conclude that if the aim of the inference is the
identification of pairs with substantial coevolution, then even simple and
computationally efficient methods like MF and PLM can provide accurate
results. If, on the contrary, model energies or probabilities have to
be accurate, then more precise methods such as ACE or BM are required.
This is for example the case if  one wants to sample from the
inferred distribution. 

\begin{figure}[h!]
   \centering
   \includegraphics[width=0.5\textwidth]{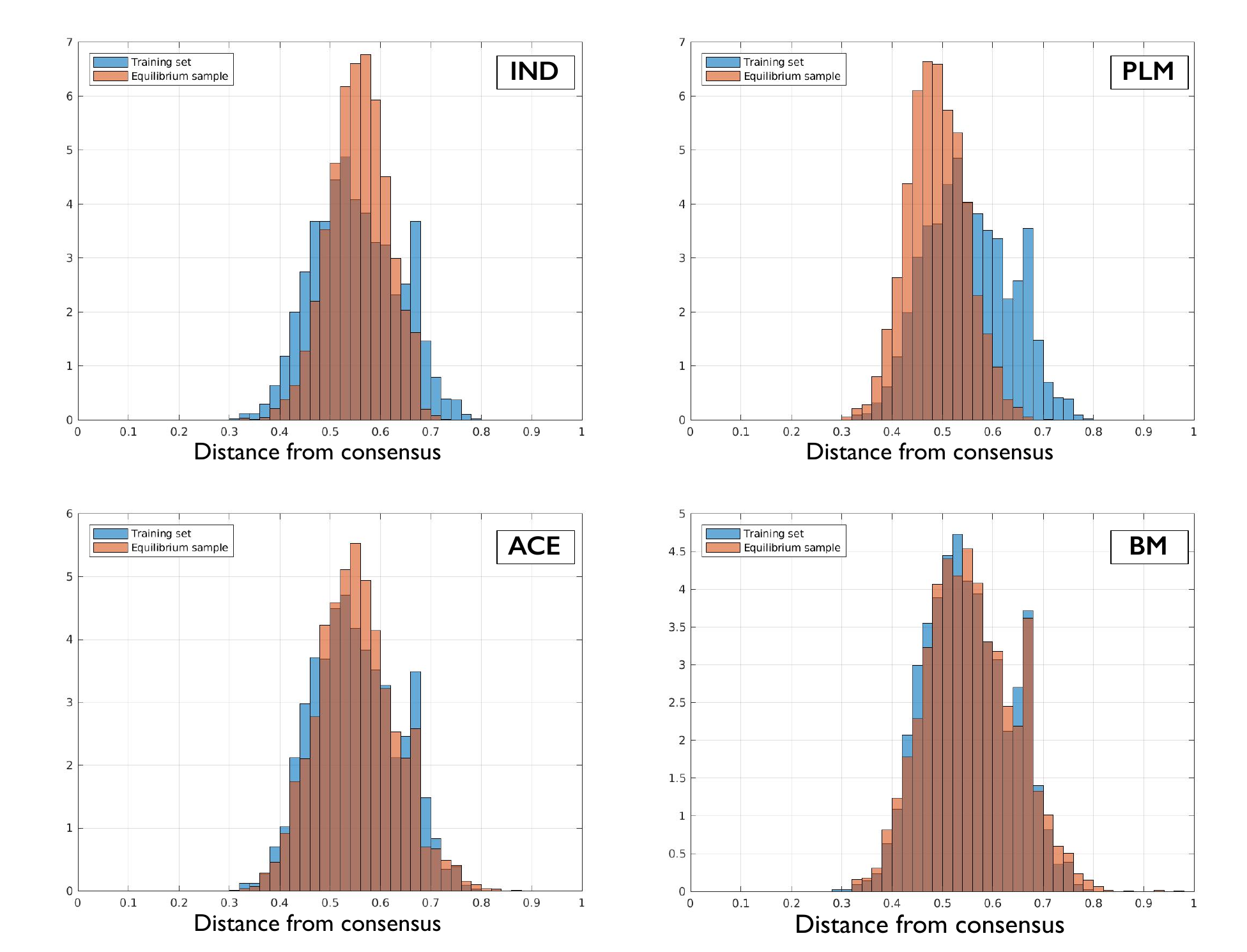}
   \caption{Comparison of different methods: The panels show
     histograms of Hamming distances of natural and model-generated
     sequences from the consensus sequence. The blue curves show the
     histograms for natural sequences (sampled from the initial MSA
     with frequencies proportional to the sequence weights), the
     orange ones for the different models (IND, PLM, ACE, BM). While
     IND and PLM show significant deviations, ACE and BM accurately
     reproduce the empirical histograms.
     Note that the measured distance is not fitted by the Potts model
     and therefore not automatically reproduced even by an accurately
     inferred Potts model.}
\label{fig:consensus}
\end{figure}

Another interesting observation is reported in
Fig.~\ref{fig:consensus}, which shows histograms of Hamming
distances of sequences to the consensus sequence
$(a_1^\star,...,a_L^\star)$ given by the most frequent amino acids
$a_i^\star = {\rm arg max}_a f_i(a)$  in each column $i$, comparing
natural sequences from the original MSA to sequences sampled from
different inferred models. While the independent model IND
reproduces the average distance (by definition of IND and the linearity of
the Hamming distance), the histogram is visibly more concentrated
than for natural sequences. PLM shows systematic deviations due to the
lack of fitting accuracy, while ACE and BM accurately reproduce the
empirical histograms. This is astonishing since the histogram measures
quantities which are {\it not} fitted via the Potts model, and
illustrates the capacity of statistical models with pairwise couplings
to well capture the sequence variability even beyond pairwise
observables. 

\subsection{Residue-residue contact prediction and tertiary structure}

The original motivation underlying the use of Potts models for describing the sequence
variability across evolutionary related proteins is the prediction of
residue-residue contacts \cite{weigt2009identification}. This task is considered hard
but important  in bioinformatics: Experimental determination of protein
structures (X-ray crystallography, NMR, cryo-electron microscopy) is expensive,
time consuming, and frequently of uncertain outcome. On the other hand, freely
accessible sequence databases are growing fast. If it were possible to use such
sequence data for predicting contacts between residues, this information could
in turn help to predict protein structures (intra-protein contacts) or to
assemble multi-protein complexes (inter-protein contacts). Since protein
function typically relies on protein structure (e.g. via binding of other
molecules to well defined interfaces on the protein surface), this would
provide crucial information about the operation of the proteins and  the
biological processes they participate in.  The validity of contact predictions
is assessed by comparison with known structures of proteins belonging to the
family under consideration. This structural information is complementary to the
sequence data and only available for a fraction of protein families.

Within DCA, residue position pairs $(i,j)$ are sorted according to
their interaction strengths as measured by $F_{ij}^{APC}$. We expect
the most coupled pairs to be in contact. High scores are thought to
indicate compensatory mutations in neighbouring residues in a
protein structure (cf. Introduction).

Pairs of residues $i,j$ with short separation $|i-j|$ along the
backbone frequently possess large 
$F_{ij}^{APC}$-values. These large scores often come from the
presence of  stretches of gaps, and are not very much relevant as far
as structural predictions are concerned.  
For this reasons and since long-range (along the protein backbone)
contacts are more informative about the tertiary structure 
most predictors do not include pairs with $|i-j|\leq 4$ into the evaluation, a
distance corresponding to one turn in an $\alpha$-helix.

\begin{figure}[h!]
          \centering
          \includegraphics[width=0.5\textwidth]{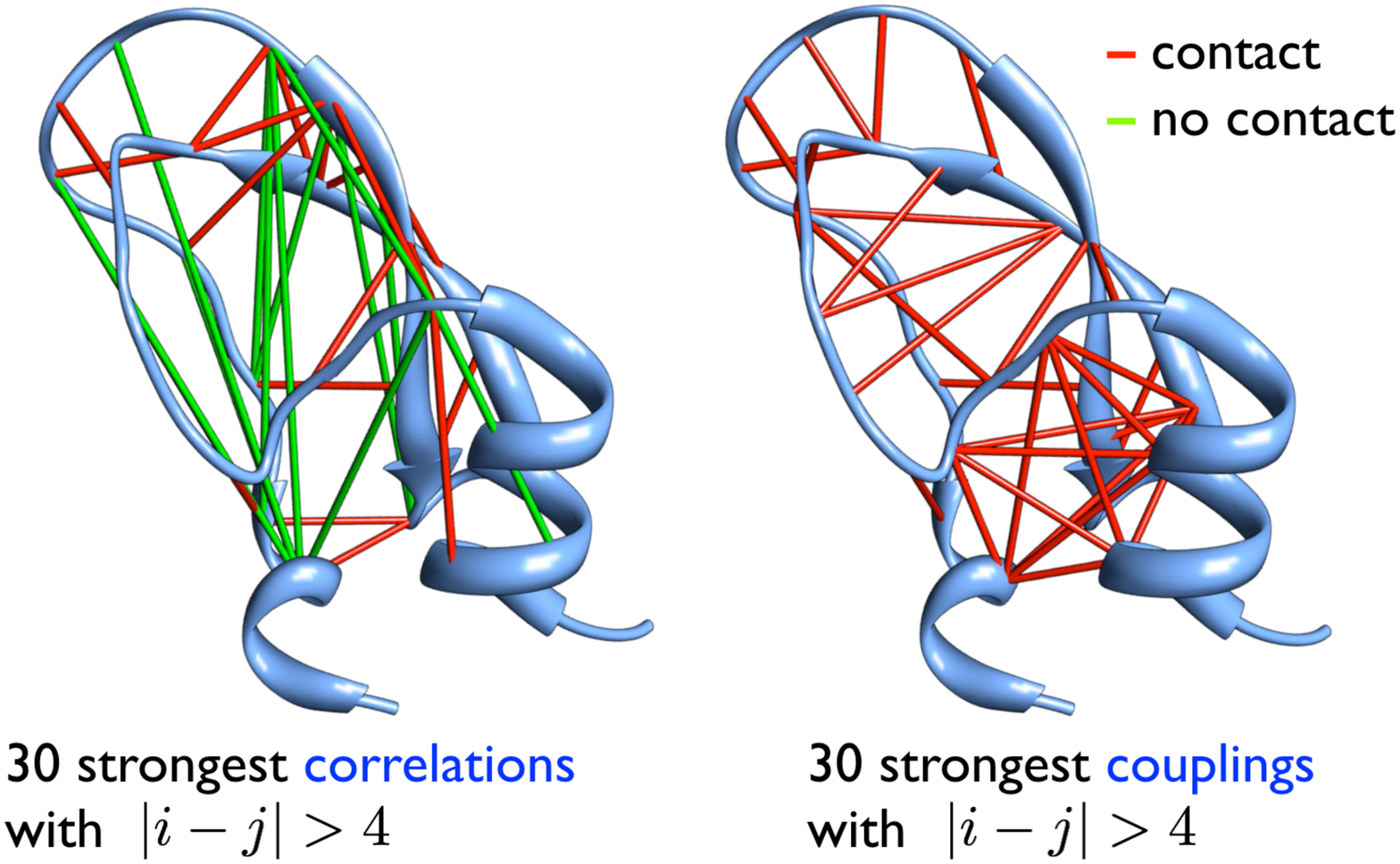}
          \caption{Residue-residue contact prediction
            based on ranking of correlations (left) and Potts  couplings (right) for the PF00014 protein
          family (trypsin inhibitor). Top    30 values of the mutual information (i.e. correlations)
          and $F^{APC}$ (i.e. couplings) scores  are mapped onto
          the      protein structure (PDB ID: 5PTI), excluding trivial 
          predictions with separation $|i-j|\leq 4$ along the
          amino-acid sequence. Red bars correspond to residue pairs in
          contact (distance below 8\AA, true positives), green bars to distant residue
          pairs (false positives). Inference was done
          using the mean-field approximation.}
 \label{fig:pf00014_contacts}
  \end{figure}

Fig.~\ref{fig:pf00014_contacts} shows an example for this procedure. Results
obtained for 4,915 sequences of length 53 amino acids, all belonging to the
Pfam protein family PF00014 (trypsin inhibitor), are mapped onto the
corresponding X-ray crystal structure (PDB ID 5PTI
\cite{wlodawer1984structure}). The left panel shows that 
the prediction by the strongest {\it correlations} (as measured by mutual
information (\ref{eq:MI})) includes a large fraction of false positives (green
lines). Predictions based on the strongest {\it couplings}, shown in the right
panel, lead to a significantly better prediction (red lines). 

While here we present only one example, the predictive performance has
been tested for thousands of protein families
\cite{morcos2011direct,ovchinnikov2017protein}. A consistent
improvement of DCA over 
correlations has been shown. Interestingly, the quality of the inference has
only a small impact on the contact prediction. The reason is that in order to
obtain a good performance in contact prediction, it is sufficient to infer the
topological aspects of the network reasonably well. While approximate methods
like MF or PLM may have a larger error in the inference of the exact parameter
values than more exact methods like BM or ACE, they robustly predict network
topology, cf.~the last section.  In fact, PLM is currently the best
stand-alone method for efficient contact prediction. 

The central use for the residue contacts extracted from MSAs with DCA is in the
prediction of protein structures. While the structure of a protein is usually
defined by its amino-acid sequence (Anfinsen's principle \cite{anfinsen1975experimental}),
the general process of protein folding \textit{in vivo} is not known. Setting
this question aside for the moment, one can also ask the easier questions on
how to computationally infer the folded structure of a protein from its
sequence. In the last decades, this was a central question in computational
biology and many methods have been devised. Even though the performance of
these methods has made impressive advances, the problem is still considered as
unsolved in general \cite{dill2012protein}. 

In this context, DCA has been shown to provide valuable information. The
predicted residues can be used as prior information on the structures. Such
information restricts the number of possible structures and makes the task thus
easier. The importance of predicted contacts in the field can be illustrated by
the fact that many of the top-performing groups in the last CASP challenge
(http://predictioncenter.org/) have made use of them. Thousands of
novel protein structures have been predicted using coevolutionary
contact predictions \cite{ovchinnikov2017protein}. 

As a last point, we note that the best performing methods for predicting
residue contacts are \textit{meta-methods}. These methods use other methods
like DCA as input and combine them using machine-learning techniques for an
improved prediction
\cite{skwark2014improved,jones2015metapsicov,wang2017accurate}. Since
they use supervised learning techniques to model the 
input-output relation between MSA and contact map, it is not surprising that
they outperform any unsupervised method in this task. 

\subsection{Predicting interaction partners and inter-protein residue
  contacts in protein-protein interaction}

The formalism of Potts Models and DCA can be used to extract biological
information on different scales. While we focused above on how to infer
contacts between residues within a protein, we now describe how to use the same
formalism to infer interactions \textit{between} proteins
\cite{ovchinnikov2014robust, feinauer2016inter}. When extended to all
proteins in a genome, this can then be used to infer an organism-wide
protein-protein interaction network.

The model in Eq.~(\ref{eq:potts}) defines a probability $P({\mathbf a})$ for finding
sequence ${\mathbf  a}$ for a specific protein belonging to protein family $A$ in an
organism. We now define a probability $P({\mathbf  a},{\mathbf b})$
for finding sequence ${\mathbf  a}$  for a
protein belonging to family $A$ and sequence ${\mathbf  b}$ for a
protein belonging to family $B$ in the 
same organism. The general idea is that if the members of the two protein
families interact in all or most organisms, they have to be mutually
compatible. Therefore, one would expect $P({\mathbf  a},{\mathbf  b})
\neq P({\mathbf  a})P({\mathbf  b})$ in such cases. 

An intuitive extension of Eq.~(\ref{eq:potts}) for two protein
sequences ${\mathbf  a}$ and ${\mathbf  b}$  is
\begin{equation*}
{\cal H}({\mathbf  a},{\mathbf  b}) = {\cal H}^{A}({\mathbf  a}) +
{\cal H}^{B}({\mathbf  b}) + {\cal H}^{AB}({\mathbf  a},{\mathbf  b}), 
\end{equation*} 
where ${\cal H}^A({\mathbf  a})$ and ${\cal H}^B({\mathbf  b})$ are of
the form of Eq.~(\ref{eq:potts}) 
and model the coevolution between residues \textit{within} the proteins $A$ and
$B$. The additional interaction term 
\begin{equation*}
{\cal H}^{AB}({\mathbf  a},{\mathbf  b})
 =  - \sum\limits_{i=1}^{L_a} \sum\limits_{j=1}^{L_b} J_{ij}(a_i,b_j)
\end{equation*} 
contains $L_a L_b q^2$ coupling parameters, with $L_a$ and $L_b$ being the
length of the sequences ${\mathbf  a}$ and ${\mathbf  b}$. These
parameters describe the coevolution 
between residue pairs in {\it different} proteins.

The model including the interaction term defines a probability for a pair of
sequences in the same organism.  As data for the inference process we therefore
need two MSAs, one for protein family $A$ and one for protein family $B$. We
then search for pairs of sequences (one from $A$ and one from $B$) that come from
the same organism and treat their concatenation as a single
configuration of the composed Potts model.

A problem can arise when there are several sequences belonging to the same
protein family inside a single organism (such sequences are called
\textit{paralogs}). The problem which sequences to pair in such cases is called
\textit{matching} and can be solved either by biological prior information
\cite{ovchinnikov2014robust,feinauer2016inter} or by another layer of
probabilistic modeling \cite{gueudre2016simultaneous,bitbol2016inferring}. 

As an interaction score for residue pairs between two proteins the
$F^{APC}$ scores of Eq.~(\ref{eq:apc}) can be used. A possible interaction score
for two proteins is then the average of the $n$ largest $F^{APC}$ scores
between the proteins. It has been shown that $n=4$ gives a good
performance, since it takes into account the strongest signals, but
averages over a few pairs to be less susceptible to noise.

\begin{figure}[htb]
\includegraphics[width=0.3\textwidth]{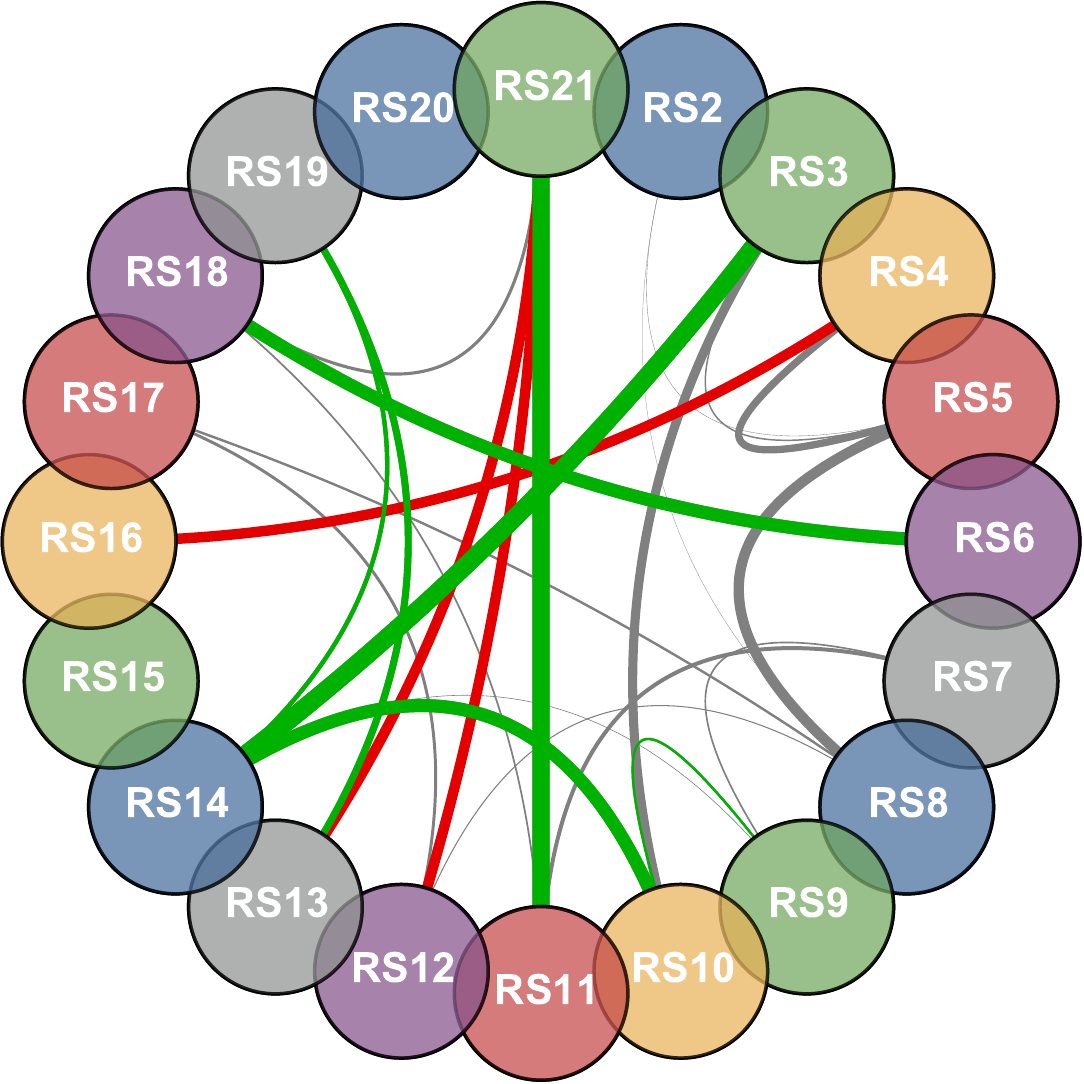}
\includegraphics[width=0.34\textwidth]{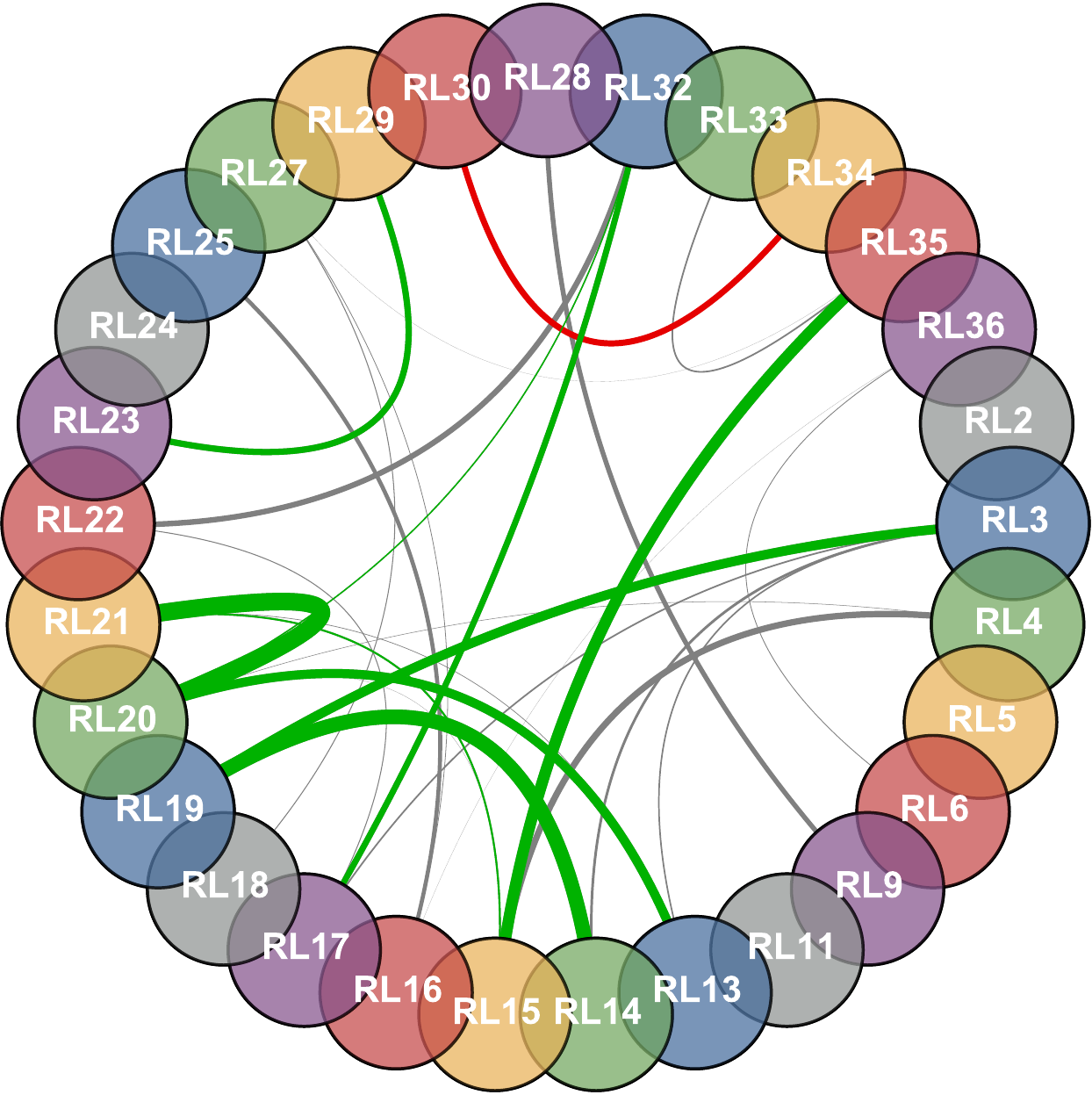}
\caption{Protein-protein interaction network of the small ribosomal subunit
(SRU, top) and the large ribosomal subunit (LRU, bottom). Circles represent
proteins, grey and green lines represent interactions found in the experimental
structure (PDBs: 2Z4K/2Z4L).  The 10 colored lines represent the 10 strongest
interaction scores according to DCA. Green indicates a true positive, red a
false positive. The width of the lines is proportional to the number of residue
contacts between the proteins (figures from \cite{feinauer2016inter}).}
\label{fig:ribo}
\end{figure}

In Fig.~\ref{fig:ribo} we represent as an example the inference
results for the two ribosomal subunits. The units consists of 49
proteins and within the experimentally determined structure (PDB:
2Z4K, 2Z4L \cite{borovinskaya2007structural}) 50 pairs are
interacting, i.e.~only 8.4\% out of the 596 intra-unit protein
pairs. We use the approach outlined above to fit a Potts Model to each
of the corresponding 596 protein-family pairs and calculated
interaction scores with $n=4$. Of the ten pairs with the largest
interaction scores in each unit, 16 are interacting in the
structure, covering most pairs with large interaction interfaces
(i.e.~fat lines in the figure). This predictive precision of 80\% has
to be compared with the 8.4\% to be expected on average in a random
prediction, i.e.~only 1.68 interacting pairs would have been found on
average when extracting 20 random pairs, with no preference for large
interfaces. 

Since one Potts Model has to be inferred per protein pair, the method is
computationally expensive for large sets of proteins. It has nonetheless been
applied to large-scale datasets, such as all protein pairs within an operon (a
set of co-localized genes expressed together) in a genome
\cite{ovchinnikov2014robust}.

The same method can also be used to infer residue-residue contacts between
proteins. Such predictions are useful when searching for the structure of a
protein complex when the structures of the single proteins are known
\cite{schug2009high,dago2012structural,ovchinnikov2014robust,
  hopf2014sequence,uguzzoni2017large}. It has been seen that large
protein-protein interfaces, which are widely conserved across species,
show reliable coevolutionary signals, while smaller interfaces or
those being only conserved in part of a protein family cannot be
easily detected by a global statistical modeling of a protein family
\cite{uguzzoni2017large}.

\subsection{Potts model for scoring: From single mutations to  entire 
  sequences} 

The Potts model inferred from the MSA of a protein family defines a
distribution over all sequences, assigning higher probabilities to sequences
likely to belong to this family.  It can thus be used to {\it quantitatively} predict
whether a sequence is similar in structure and function to the sequences the
model has been trained on. In this way the model becomes a tool for assessing
the effect of mutations in protein sequences, for predicting whether synthetic
sequences fold into a known structure or for generating new synthetic sequences
by sampling from the model.

A natural quantity to score sequences in these applications is their energy.
By definition, Eq.~(\ref{eq:potts}) is minus the log-probability of the
sequence (up to an additive constant coming from normalization). We now show
that this score leads to very promising results when tested on experimental
data.

A {\it local} test checks if energy differences (i.e. differences in
log-probability of the Potts model) actually can quantify the {\it
  fitness effect} of mutations. This has been done successfully in a
number of situations, ranging from viral over bacterial to human
proteins \cite{chakraborty2014hiv,morcos2014coevolutionary,
figliuzzi2015coevolutionary,hopf2017mutation,feinauer2017context}. 
Such predictions are of great 
biomedical interest, since they help to find mutations related to
virulence or drug resistance of pathogens to the identification of
disease-causing mutations in between the multitude of neutral
polymorphisms observable in human.
 
On a more {\it global} level, the energy of the Potts model can be
seen as a measure in how far a new (e.g. artificially designed)
amino-acid sequence is compatible with the natural sequence
variability extracted via our Potts model from known sequences (i.e. the
MSA used to infer the model parameters). This was first done in
\cite{balakrishnan2011learning} using PLM-based inference and
experimental data from
\cite{socolich2005evolutionary,russ2005natural}. The latter works are
based on a method called Statistical Couplings Analysis (SCA), which
was used for generating computationally non-natural members of the WW
protein domain family (PFAM family PF00397, with an MSA of currently
$M=12742$ sequences).   

The SCA sampling process is based only on the statistical properties
of an MSA of natural members of the family; the resulting non-natural
sequences were expressed and tested experimentally.  The properties
determined were thermodynamic stability in
\cite{socolich2005evolutionary}, and the peptide binding specificity
for a subset of stable proteins in \cite{russ2005natural}. 

The starting point for the generation of artifical sequences was a small and
curated alignment of 120 natural sequences of the WW domain family (length 35
residues), 42 of which were included in the experiments (called NAT --
natural). To generate artificial sequences, the alignment was shuffled in three
different ways: 
\begin{itemize}
\item (R -- random): A random permutation was applied to all entries in the
  matrix, thereby destroying any pattern of site-specific conservation
  (i.e. ``magnetization'') or covariation between sites. 19 sequences
  were tested experimentally.
\item (IC -- independent conservation): Sequences were obtained
  by shuffling each column of the original MSA. Statistical features
  of individual positions are therefore unchanged, while covariation
  patterns are destroyed. The IC shuffling procedure corresponds to a
  sampling from the independent-site model Eq.~(\ref{eq:para}). 43
  sequences were tested experimentally.
\item (CC -- coupled conservation): Starting from the IC data set, a
  Monte-Carlo annealing was used to to approximately reproduce
  the pairwise amino acid frequencies $f_{ij}(a,b)$ of the original
  MSA of natural sequences. The CC dataset is closely related to a
  sampling from the Potts model in Eq.~(\ref{eq:potts}),
  cf.~\cite{bialek2007rediscovering}. 43 sequences were tested
  experimentally. 
\end{itemize}
Denaturation experiments showed that none of the R or IC sequences
folded into the correct structure.  On the contrary, $31\%$ of the CC
sequences and $67\%$  of the NAT sequences folded correctly in the
experimental conditions used.  These results are a strong indication
that limited sequence information might be sufficient for defining the
structural constraints acting on the evolution of the WW domain 
family. Considering the complexity of interactions between amino acids,
illustrated by the notorious difficulty of solving  the protein folding
problem \cite{dill2012protein}, this may come as a surprise.
More exactly, reproducing the pairwise correlations in the
amino-acid distribution, in addition to the single-site frequencies,
seems to be necessary and (almost) sufficient to specify the native fold.

However, the SCA algorithm was not able to score {\it single}
sequences, but only the entire shuffled alignement. On the contrary,
single sequences can be assessed by the Potts-model energies described
before. It was first noted in \cite{balakrishnan2011learning}, that
these energies are effectively an excellent predictor which of the
sequences fold, and which not.

\begin{figure}[htb!]
          \centering
          \includegraphics[width=0.5\textwidth]{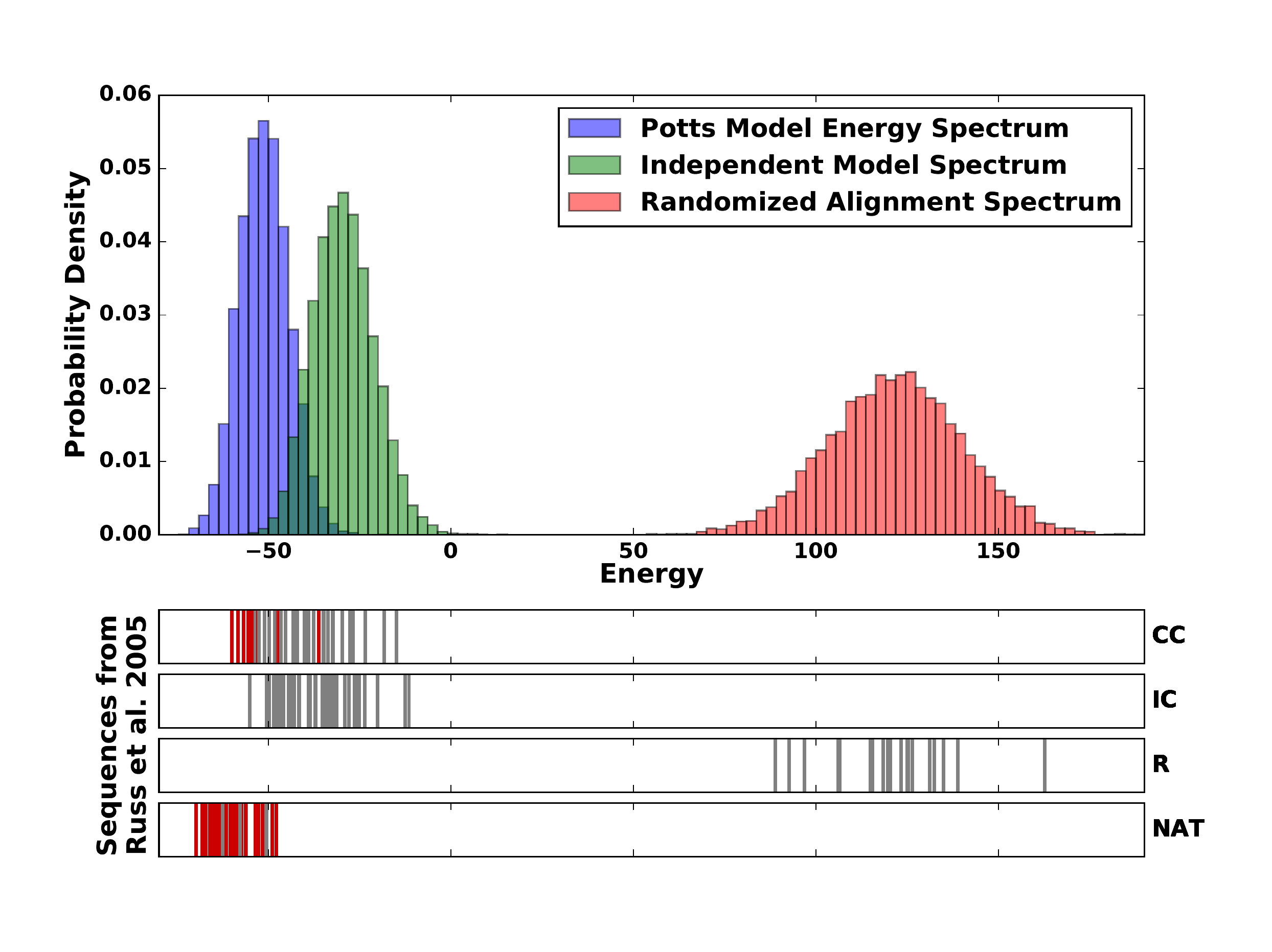}
          \caption{ WW sequences: Potts energies and folding qualities. Top:  Distribution of  Potts energies with couplings and fields parameter inferred by the ACE algorithm sampled by MCMC  from the coupled Potts (red) and the
          Independent-site (green) models. Bottom: First row: Energies of the  43 CC sequences, among which 12 well folded  (red) and 31 (gray) not well folded.   Second row: Energy of the  43 IC sequences, 
          among which none of them fold. Third rows: Energies of the  42 Natural Sequences, among which   28 well folded (red) and 14 (gray) not well folded according to the denaturation tests. The 19 random sequences have all higher energies ${\cal H}>0$.  }
 \label{fig:foldingWW}
  \end{figure}

Using the couplings $J_{ij}(a,b)$ and fields $h_i(a)$ inferred with a
recent Pfam release, sequence energies for the different sets were
calculated. We compare them to energies of sequences sampled from the same
Potts model and the independent model in Fig.~\ref{fig:foldingWW}. In the top
panel, the energy distribution of sequences generated from a Monte Carlo
sampling of the inferred Potts model (blue) is compared to the distribution of
energies of sequences generated by the independent model (green) and
from random sequences (red). In the bottom panel, the sequences tested
experimentally in \cite{socolich2005evolutionary} are shown - by red
bars if they were folding in the experiments, by grey bars if not. 

As a first observation, we see that also the IC sequences (as well as
the sequences sampled from the independent model) have significantly
smaller energies than the random sequences, even if these are
evaluated with the pairwise Potts model. The energy spectra of the
sample from the Potts and the independent model are partially
overlapping, resulting also in overlapping energy values of the CC and
IC sequences. The most striking observation is, however, that most
folding CC and N sequences have energies smaller than those achieved
by IC and independent-model sequences, while the CC and N sequences
with higher energies, in the range of IC energies, are mostly non folding. The
energy turns out to be an excellent discriminator between folding and
not folding sequences across the NAT, CC, IC and R data sets.

Fig.~\ref{fig:foldingWW} shows the results obtained using ACE
for model inference. While the other inference techniques (MF, PLM,
BM) show some quantitative differences, their performance in
discriminating folding from non folding sequences is very comparable:
The performance of Potts models in such classification tasks does not depend
much on the inference technique used. This is probably due to the fact
that the inferred parameters have to be only good enough to rank the sequences
correctly, such that a more precise inference does not lead to a better
performance in sequence ranking.


\subsection{Generative aspects and entropy: from lattice proteins to HIV}

An ambitious goal is to generate new and functional protein sequences, e.g. by
Monte Carlo sampling from the inferred model, along the lines of the study of the WW domain by Ranganathan and collaborators \cite{socolich2005evolutionary,russ2005natural}, cf.~Fig.~\ref{fig:foldingWW}.
The question was also addressed in \cite{jacquin2016benchmarking} in
the highly idealized case of lattice proteins, a model of 27
amino-acid long chains folding on discrete $3\times3\times3$  cubic structures
\cite{shakhnovich1990enumeration}.  In this setting, protein
families are defined as the set of sequences folding into one of the $\sim10^5$
structures on the cube.  Jacquin et al.~\cite{jacquin2016benchmarking}
found that new sequences generated by MCMC from the ACE-inferred Potts
model describing a structural family have high probability to fold into the same structure (but not for less precise Potts models
based on MF and PLM inference). On the contrary, sequences
sampled from the independent model rarely fold. These
results confirm -- in the simple case of lattice proteins -- the claim
of \cite{socolich2005evolutionary,russ2005natural} that keeping 2-point
statistical information is necessary and sufficient for generating structurally valid
proteins.

As discussed above, the cross-entropy is an estimate of the Gibbs-Shannon
entropy of the sequence distribution of proteins belonging to the same family,
based on the limited sample of known functional sequences.  Roughly speaking,
the entropy can be thought of as the logarithm of the number of sequences in
the family. This definition is approximate, as some sequences may express the biological
function with varying degrees. Computing the value of the entropy is of interest,
since it allows us to quantify the diversity of
possible proteins sharing a common biological function. The size of a protein
family is expected to be much larger than the size of the available MSA.
Considering again lattice proteins on the  $3\times3\times3$  cube allows one to obtain a
quantitative understanding of thses concepts in an idealized case
\cite{jacquin2016benchmarking}. Calculations show that the $\sim10^5$
families defined by the possible structures contain each a variable
number of protein sequences.  The numbers range between $10^{20}$ and
$10^{25}$ \cite{barton2016entropy} and depend on the designability of the
native fold \cite{li2002designability,england2003structural}. The
total number of sequences in any of the structures represents thus a tiny
fraction of the total number of sequences, $20^{27}\simeq 10^{35}$.  

While estimating the entropy of real protein families is a daunting,
not well-defined task, approximate results can be obtained from the
crossentropy when using ACE to infer Potts models
\cite{barton2016entropy}. For the case of the 
WW domain discussed before, this amounts to approximately 1.2 nats
per residue position, i.e. to about
$e^{1.2}\simeq 3.3$ different amino acids per site.  As the Potts model
takes into account only the one- and two-point statistics, we expect the true
entropy, which reflects higher order statistical
constraints, to be smaller (note that the independent-site model,
which does not take into account pairwise correlations,  gives  
a higher entropy density of about $\ln 5.5\simeq 1.7$ nats).
All these estimates are obviously much smaller than
the value $\ln 20 \simeq 2.99$ nats, which would be obtained for purely
random amino-acid sequences. They are closer to the value of $1.9$ nats
obtained by Shakhnovich an collaborators from purely thermodynamical
considerations \cite{shakhnovich1998protein}.

The concept of entropy was also recently used  in the study of HIV
viral sequence variability. HIV is 
characterized by a large sequence mutability:  the viral population
constantly escapes from the immune system by mutating amino acids in
the epitope, a subsequence of  about 10 amino acids, which is
recognized by antibodies. Understanding the mutational landscape of
HIV sequences is therefore of fundamental importance in the hope to
design drugs or vaccines that block as many escape mutations as 
possible. As compensatory mutations (i.e. epistasis couplings between
amino acids, as included in the Potts models of protein sequences)
play a central role in the escape mutations, it is important to use
predictive models that take into account couplings between amino
acids. In the following we illustrate a few studies, allowing
to more accurately characterize the mutational landscape.
 
\begin{figure}[h!]
          \centering
          \includegraphics[width=0.4\textwidth]{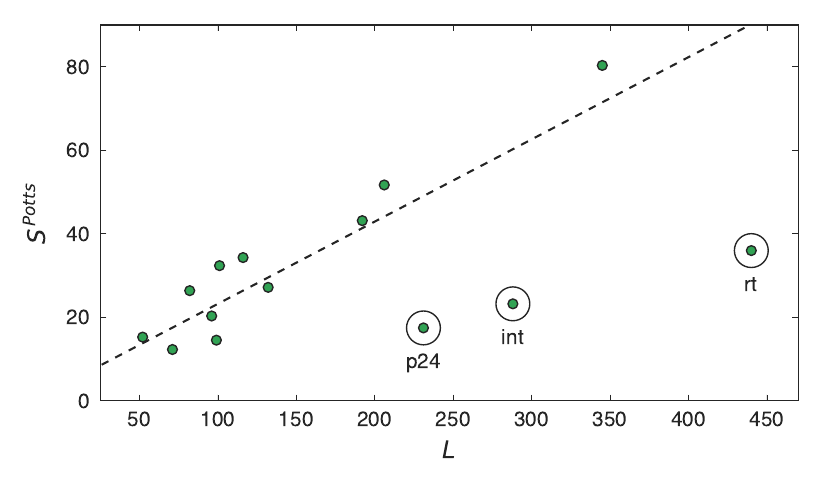}
          \caption{ Entropies of the 14 protein families of the HIV virus, as
            estimated by the ACE algorithm extracted from the
            alignments (Los Alamos data base), vs. length of the
            protein. Results are shown for Clade  B of the HIV virus,
            and are similar to what is found for Clade C, see
            \cite{barton2016entropy}. }  
 \label{fig:entropy14hiv}
  \end{figure}

In Fig.~\ref{fig:entropy14hiv}, we plot the cross entropies of all the 14
proteins encoded in the HIV virus genome. These quantities allow us to identify
the viral proteins that are more conserved and therefore less inclined to
mutations \cite{barton2016entropy}. Interestingly, among the three more conserved
proteins, the reverse transcriptase and integrase are not frequently targeted
by the immune system, and therefore not under its selective pressure. On the
contrary,  protein p24, which forms the viral capsid, is frequently targeted by
the host immune system. Its large conservation despite this immune pressure
suggests that this protein is tightly constrained. Epitopes in this region have
been shown to be frequently targeted by individuals that efficiently control
the viral infection \cite{dahirel2011coordinate}.  In addition one can predict to
what extent variations in amino acids at individual sites contribute to the
total entropy of the protein. Neglecting couplings between sites, this can be
easily obtained from data \cite{ferrari2011relationship} using the single site
entropy as estimated from the empirical frequencies of amino acids on the
sites, $S_{site}(i) =-\sum_a f_i(a) \log f_i(a)$.  Corrections to these
estimates can be done once the Potts couplings are inferred, as shown in
\cite{barton2016entropy}.
    
More results were obtained by applying the Potts approach to HIV sequence data.
In \cite{chakraborty2014hiv}, the energy cost of sequence mutations with respect to
the wild type sequence has been estimated by the inferred Potts model and
compared to in vitro measures of the viral replicative power, which is a direct
measure of this fitness. A correlation coefficient of $-0.76$ was observed. In
\cite{barton2016relative},  data consisting of samples of the viral sequences and
antibody populations in the same patients over time have been analyzed and
compared to Potts model prediction.  It has been shown that two patients, in
which the same epitope is targeted in the same protein, have escaping times
that are very different. Such observations are explained by the fact that the
overall viral sequence is different in the two patients and therefore the cost
of the escaping mutations, according to the energy cost estimated by the Potts
model, is different due to the coupling terms. The  energy cost is therefore a
good estimator  for the time for the escaping mutations .



\section{Conclusion and outlook}

In this review, we have illustrated how methods borrowed from (inverse)
statistical physics are gaining influence in the analysis of massive
sequence data. Such data become available at an unprecedented pace
thanks to next-generation sequencing. The large amounts of data pose
both a challenge and an opportunity for sophisticated methods of data
analysis and modeling: 
\begin{itemize}
\item From the point of view of {\em bioinformatics}, computational
  methods are needed to analyse data and gain insight into specific
  systems of biological and biomedical interest. An example is the use
  of sequence data to predict the three-dimensional structure of a
  folded protein.
\item From the point of view of {\em statistical biological physics},
  such large amounts of data help to gain deep insight into general
  principles governing biological systems and their evolution. As an
  example, analysing the sequence variability between evolutionary
  related proteins provides information about how and which structural
  and functional characteristics of proteins constrain its evolution
  over millions of years.
\end{itemize}
Within the statistical modeling of large biological data sets, both
viewpoints meet and reinforce each other, placing the area of research
at the interface between statistical physics, bioinformatics and
biology. 

Despite the success of modeling protein sequences via Potts models,
there are important limitations, which require substantial theoretical
and applied work in the years to come:
\begin{itemize}
\item The success of pairwise models is astonishing -- there is no
  fundamental reason why higher-order statistical features (and thus
  higher-order statistical couplings) should not play a role. It is
  interesting to unveil the reasons of this success, as well as its
  limitations. This question is particularly important in the context
  of the discussed generative models -- a true generative model should
  produce sequences which, by no statistical means, are
  distinguishable from natural sequences. 
\item Current modeling schemes assume amino-acid sequences to be
  configurations of abstract symbols, which have no meaning on their
  own. Obviously, there is much prior knowledge about amino-acids, there
  are many methods to predict, e.g., the secondary structure of a
  protein based on its sequence, the solvent accessibilty of a
  residue, or the propensity of two residues to form a contact based
  on their physicochemical properties. So far, such complementary
  knowledge is not taken into  account. As is illustrated by
  metapredictors using machine-learning techniques for combining
  coevolutionary scores with prior knowledge, the {\em
    integration of different data sources} is of the utmost 
  importance. Systematic approaches combining heterogenous sources of 
  information in a principled way have to be developped within the
  statistical-physics community.
\item Up to empirical statistical corrections (reweighting), current
  models assume sequences to be an independently and identically
  distributed sample drawn from some unknown probability
  distribution. For sequence data, this assumption is wrong --
  biological sequences have a non-trivial phylogenetic distribution,
  and the independent evolution between two recently divided species
  is typically not ``at equilibrium'', i.e. the sequences cary memory
  about the common ancestor. What is the correct treatment of such
  sampling biases? 
\item Last but not least, the methods have been tested in general on
  sample cases with known answer. Large-scale predictions of, e.g.,
  unknown protein structure, are still rare.
\end{itemize}
In the meanwhile, databases are growing and more and more biological
systems become amenable to statistical-physics inspired
approaches. Similar approaches have been applied to
systems as different as gene-regulatory networks
\cite{lezon2006using,margolin2006aracne,bailly2010inference},  retinal
and hippocampal neural spiking data 
\cite{schneidman2006weak,tkacik2006ising,cocco2009neuronal,posani2016}
and the collective behavior of animal groups
\cite{bialek2012statistical,cavagna2014dynamical}. We are certain 
that more systems will be added to this list.

\section*{Acknowledgements}

SC, RM and MW acknowledge funding by the ANR project COEVSTAT
(ANR-13-BS04-0012-01). This work was undertaken partially (MF and MW) in
the framework of CALSIMLAB, supported by the grant ANR-11-LABX-0037-01
as part of the "Investissements d'Avenir" program
(ANR-11-IDEX-0004-02). We thank J.P. Barton for  his help on
Fig.~\ref{fig:pf00014_contacts}.

\bibliography{review} 

\end{document}